\RequirePackage[displaymath, mathlines]{lineno}
\documentclass[aps,prd,superscriptaddress,showpacs,floatfix,twocolumn]{revtex4-1}

\usepackage{graphicx}
\usepackage{hyperref}
\usepackage{amsmath}
\usepackage{breakurl}
\usepackage[caption=false]{subfig}

\usepackage{color}

\usepackage{times}
\begin{document}


\title{Lepton angular distributions of fixed-target Drell-Yan experiments in
  perturbative QCD and a geometric approach}

\author{Wen-Chen Chang}
\affiliation{Institute of Physics, Academia Sinica, Taipei 11529, Taiwan}

\author{Randall Evan McClellan}
\affiliation{Thomas Jefferson National Accelerator Facility,
Newport News, VA 23606, USA}
\affiliation{Department of Physics, University of Illinois at
Urbana-Champaign, Urbana, Illinois 61801, USA}

\author{Jen-Chieh Peng}
\affiliation{Department of Physics, University of Illinois at
Urbana-Champaign, Urbana, Illinois 61801, USA}

\author{Oleg Teryaev}
\affiliation{Bogoliubov Laboratory of Theoretical Physics,
JINR, 141980 Dubna, Russia}

\date{\today}

\begin{abstract}

The lepton angular distributions of the Drell-Yan process in
fixed-target experiments are investigated by NLO and NNLO perturbative
QCD. We present the calculated angular parameters $\lambda$, $\mu$,
$\nu$ and the degree of violation of the Lam-Tung relation,
$1-\lambda-2\nu$, for the NA10, E615 and E866 experiments. Predictions
for the ongoing COMPASS and SeaQuest experiments are also
presented. The transverse momentum ($q_T$) distributions of $\lambda$
and $\nu$ show a clear dependence on the dimuon mass ($Q$) while those
of $\mu$ have a strong rapidity ($x_F$) dependence. Furthermore,
$\lambda$ and $\nu$ are found to scale with $q_T/Q$. These salient
features could be qualitatively understood by a geometric approach
where the lepton angular distribution parameters are expressed in
terms of the polar and azimuthal angles of the ``natural axis'' in the
dilepton rest frame.

\end{abstract}

\maketitle

\section{Introduction}
\label{sec:introduction}

The Drell-Yan (D-Y) process~\cite{drell} is one of the important
experimental approaches to explore the partonic structure of
hadrons~\cite{peng14}. It is a unique tool for accessing the
structures of unstable hadrons such as pions and
kaons~\cite{falciano86,conway,dutta2013}. The D-Y process plays an
essential role in probing the sea quarks of
protons~\cite{NA51,e866,chang14} as well. The transverse momentum
($q_T$) distributions of the D-Y cross sections yield important
information on the intrinsic transverse momentum ($k_T$) distribution
of partons~\cite{Bacchetta:2017gcc} in the small-$q_T$
region. Furthermore, the polar and azimuthal angular distributions of
leptons produced in unpolarized D-Y process are sensitive to the
underlying reaction mechanisms and to novel parton distributions such
as Boer-Mulders functions~\cite{boer99}. For measurement with a
transversely polarized target, a recent experiment extracted
information on Sivers functions for the first time via the D-Y
process~\cite{compass}.

In the rest frame of the virtual photon in the D-Y process, a commonly
used expression for the lepton angular distributions is given
as~\cite{lam78}
\begin{equation}
\frac{d\sigma}{d\Omega} \propto 1+ \lambda \cos^2\theta
+ \mu \sin 2 \theta\cos\phi
+ \frac{\nu}{2} \sin^2\theta \cos 2 \phi,
\label{eq:eq1}
\end{equation}
where $\theta$ and $\phi$ refer to the polar and azimuthal angles of
$l^-$ ($e^-$ or $\mu^-$). At leading-order (LO),
$q\bar{q}\rightarrow \gamma^*$ with collinear partons leads to a
transversely polarized virtual photon with a prediction of $\lambda =
1$ and $\mu = \nu =0$~\cite{drell}. To describe the D-Y process with
finite $q_T$, higher-order QCD processes, such as $q\bar{q}\rightarrow
\gamma^* G$ and $qG \rightarrow \gamma^* q$ in
$\mathcal{O}(\alpha_S)$, should be included and these processes could
alter the angular coefficients $\lambda$, $\mu$ and $\nu$ in
principle. While $\lambda$ can now deviate from 1, and $\mu$ and $\nu$
can be nonzero, a well-known result is that the Lam-Tung (L-T)
relation~\cite{lam80},
\begin{equation}
1-\lambda-2\nu=0,
\label{eq:eq2}
\end{equation}
holds for both NLO processes. Deviation from the L-T relation appears
in the NNLO process $\mathcal{O}(\alpha_S^2)$ and beyond, e.g.
$q\bar{q}\rightarrow \gamma^* GG$, $qG \rightarrow \gamma^* qG$ and
$GG \rightarrow \gamma^* G$ according to
pQCD~\cite{Brandenburg:1993cj}.

Violation of the L-T relation was observed in the fixed-target
experiments with pion beams by NA10~\cite{falciano86} and
E615~\cite{conway}, while L-T was found to be satisfied in the D-Y
production with proton beams by E866~\cite{zhu}. The $q_T$ range of
these fixed-target experiments is between 0 and 5 GeV. As for the
measurements of $Z$ boson production in the collider experiments, CDF
data of $p-\bar{p}$ collision~\cite{cdf} are consistent with the L-T
relation, while CMS and ATLAS data of $p-p$ collision~\cite{cms,atlas}
show a clear violation. The violation of the L-T relation at $q_T>5$
GeV could be well described taking into account NNLO pQCD
effect~\cite{Gauld:2017tww}. Lambersten and
Vogelsang~\cite{Lambertsen:2016wgj} compared the NLO and NNLO pQCD
calculations of $\lambda$ and $\nu$ with the data of fixed-target
experiments NA10, E615 and E866. Overall the agreement is not as good
as seen in the collider data at large $q_T$.

Recently we interpreted the violation of the L-T relation as a
consequence of the acoplanarity of the partonic
subprocess~\cite{peng16,chang17}. This acoplanarity can arise from
intrinsic transverse momenta of partons inside the hadrons, or from
the perturbative gluon radiation beyond $\mathcal{O}(\alpha_s)$ such
that the axis of the annihilating quark-antiquark pair (natural axis) no longer necessarily resides on the colliding hadron plane. In addition to the violation of the L-T relation, other salient features
of the $q_T$ dependence of the $\lambda$, $\mu$ and $\nu$ parameters
of the $Z$ production data from the collider
experiments~\cite{peng16,chang17}, as well as the rotational
invariance properties of these parameters~\cite{peng18}, could be well
explained by this intuitive geometric approach.

In this work we compare the $\lambda$, $\mu$, $\nu$ data measured at
NA10~\cite{falciano86}, E615~\cite{conway} and E866~\cite{zhu} with
the fixed-order pQCD calculations. The approach is similar to what was
done in Ref.~\cite{Lambertsen:2016wgj}, but we extend the study to
include the L-T violation quantity $1-\lambda-2\nu$, the $\mu$
parameter, as well as the scaling behavior of these angular
parameters. Furthermore we present the NLO pQCD predictions for the
ongoing COMPASS~\cite{COMPASSII} and SeaQuest~\cite{E906} experiments
on the dimuon mass $Q$ and Feynman-$x$ ($x_F$) dependence of the
angular parameters. The common features between the pQCD and the
geometric approach~\cite{peng16,chang17} are also discussed.

This paper is organized as follows. In Sec.~\ref{sec:method}, we
describe how the fixed-order pQCD calculation is performed to extract
the angular distribution parameters. The results from the pQCD
calculations for the existing and forthcoming fixed-target experiments
are then presented in Secs.~\ref{sec:results_oldexp}
and~\ref{sec:results_newexp}, respectively. We further interpret some
notable features of pQCD results using the geometric model in
Sec.~\ref{sec:discussion}, followed by conclusion in
Sec.~\ref{sec:conclusion}.

\section{Calculations of angular parameters in DYNNLO}
\label{sec:method}

The formalism of the NLO ($\mathcal{O}(\alpha_S)$)~\cite{DY_nlo} and
the NNLO ($\mathcal{O}(\alpha_S^2)$)~\cite{DY_nnlo} QCD of the D-Y
process have been known for a while. It is not until recently that
packages of evaluating the differential D-Y cross sections up to
$\mathcal{O}(\alpha_s^2)$ from $p-p$ and $p-\bar{p}$ collisions are available
for public usage: DYNNLO~\cite{DYNNLO} and FEWZ~\cite{FEWZ}. Both
packages are parton-level Monte Carlo programs and they provide the
differential cross sections for the D-Y process and $W$/$Z$ vector
boson production.  The threshold resummation of soft-gluon emission at
small $q_T$ is not included in these two packages. As discussed in
Ref.~\cite{Lambertsen:2016wgj}, even though resummation is important
for the cross sections, it is expected not to affect the angular
parameters~\cite{boer06,berger}.

In this work we utilize the DYNNLO (version 1.5)
package~\cite{DYNNLO_Web}. With some minor modifications, the code can
evaluate the D-Y cross sections induced by pion or proton beams on
proton or neutron targets. Via the LHAPDF6 framework~\cite{LHAPDF6},
the parton distribution functions (PDFs)~\cite{PDFsets} used for the
protons and neutrons are ``CT14nlo'' and ``CT14nnlo'' in the NLO and
NNLO calculations, respectively, and ``GRVPI1'' for the pion PDFs in
both NLO and NNLO calculations. The factorization scale ($\mu_F$) and
renormalization scale ($\mu_R$) are set as $\mu_F = \mu_R = Q$.

In order to calculate the $\lambda$, $\mu$, and $\nu$ parameters, we
first calculate the $A_i$ parameters in an alternative expression of
the lepton angular distributions of the D-Y process as
follows~\cite{cs}:
\begin{eqnarray}
\frac{d\sigma}{d\Omega} & \propto & (1+\cos^2\theta)+\frac{A_0}{2}
(1-3\cos^2\theta) \nonumber \\
& & +A_1 \sin 2 \theta\cos\phi + \frac{A_2}{2} \sin^2\theta \cos 2 \phi
\label{eq:eq3}
\end{eqnarray}
where $\theta$ and $\phi$, same as in Eq.~(\ref{eq:eq1}), are the
polar and azimuthal angles of $l^-$ ($e^-$ or $\mu^-$) in the rest
frame of $\gamma^*$. The angular coefficients $A_i$ could be evaluated
by the moments of harmonic polynomial expressed
as~\cite{atlas,Gauld:2017tww}
\begin{eqnarray}
A_0 & = & 4 - 10 \langle \cos^2\theta \rangle, \nonumber \\
A_1 & = & 5 \langle \sin2\theta \cos \phi \rangle, \nonumber \\
A_2 & = & 10 \langle \sin^2\theta \cos2\phi \rangle,
\label{eq:eq4}
\end{eqnarray}
where $\langle f(\theta, \phi) \rangle$ denotes the moment of
$f(\theta, \phi)$ , i.e. the weighted average of $f(\theta, \phi)$ by
the cross sections in Eq.~(\ref{eq:eq3}). It is straightforward to
show that $\lambda, \mu, \nu$ in Eq.~(\ref{eq:eq1}) are related to
$A_0, A_1, A_2$ via
\begin{eqnarray}
\lambda = \frac{2-3A_0}{2+A_0};~~~ \mu  =  \frac{2A_1}{2+A_0};~~~
\nu  =  \frac{2A_2}{2+A_0}.
\label{eq:eq5}
\end{eqnarray}
Equation~(\ref{eq:eq5}) shows that the L-T relation, $1-\lambda - 2
\nu=0$, is equivalent to $A_0 = A_2$.

\section{Comparison with existing data from NA10, E615 and E866}
\label{sec:results_oldexp}

\begin{figure}[htbp]
\includegraphics[width=1.0\columnwidth]{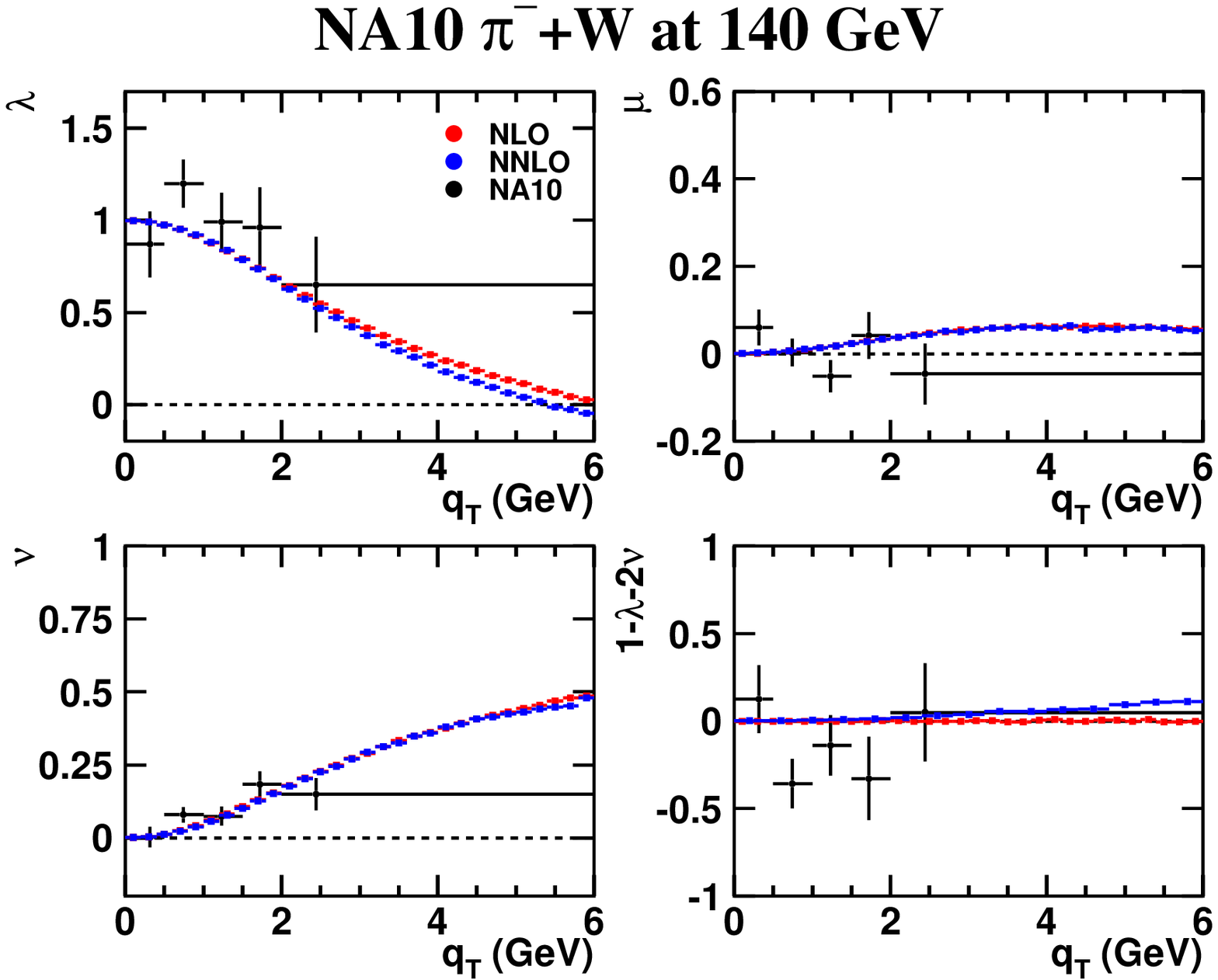}
\caption{Comparison of NLO (red points) and NNLO (blue points)
  fixed-order pQCD calculations with the NA10 $\pi^-+W$ D-Y data at 140
  GeV~\cite{falciano86} (black points) for $\lambda$, $\mu$, $\nu$ and
  $1-\lambda-2\nu$.}
\label{fig1_na10_140}
\end{figure}

\begin{figure}[htbp]
\includegraphics[width=1.0\columnwidth]{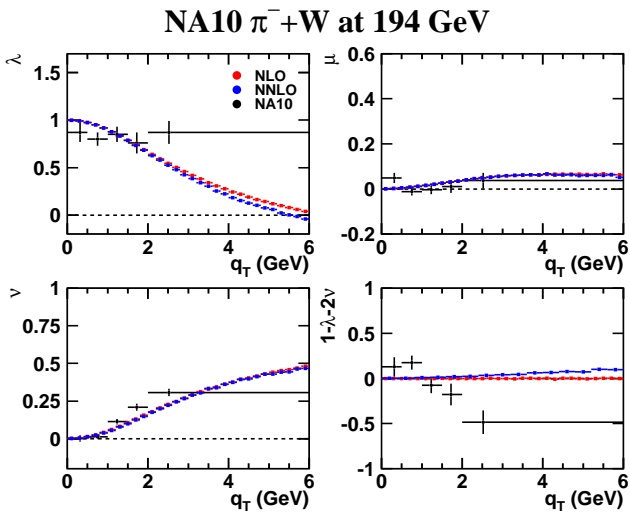}
\caption{Same as Fig.~\ref{fig1_na10_140}, but for NA10
  data~\cite{falciano86} with 194-GeV $\pi^-$ beam.}
\label{fig1_na10_194}
\end{figure}

\begin{figure}[htbp]
\includegraphics[width=1.0\columnwidth]{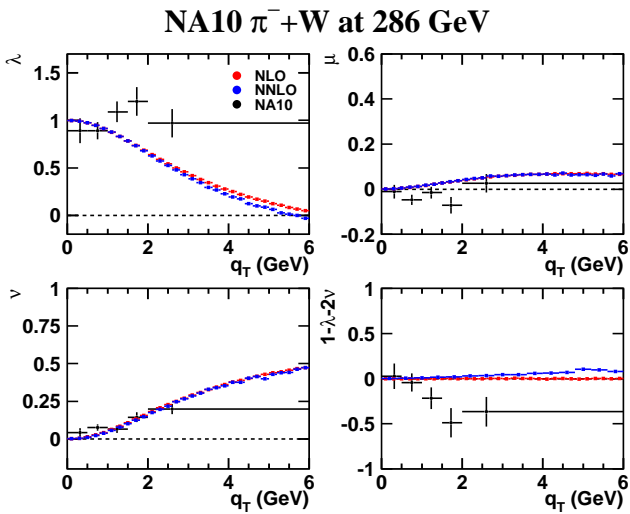}
\caption{Same as Fig.~\ref{fig1_na10_140}, but for NA10
  data~\cite{falciano86} with 286-GeV $\pi^-$ beam.}
\label{fig1_na10_286}
\end{figure}

\begin{figure}[htbp]
\includegraphics[width=1.0\columnwidth]{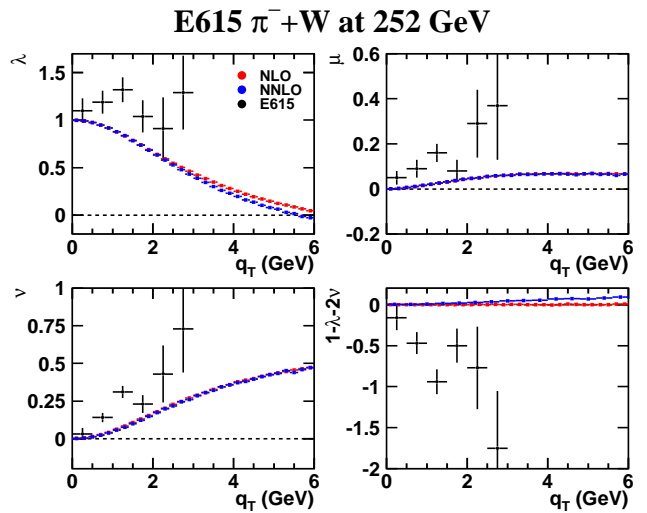}
\caption{Comparison of NLO (red points) and NNLO (blue points)
  fixed-order pQCD calculations with the E615 $\pi^-+W$ D-Y data at 252
  GeV~\cite{conway} (black points) for $\lambda$, $\mu$, $\nu$ and
  $1-\lambda-2\nu$.}
\label{fig1_e615}
\end{figure}

Now we compare the results of $\lambda$, $\mu$, $\nu$, and the L-T
violation, $1-\lambda-2\nu$, from the fixed-order pQCD calculations
with existing data from fixed-target experiments. The angular
parameters are evaluated as a function of the dimuon's $q_T$ in the
Collins-Soper frame~\cite{cs}. We first consider the data from
NA10~\cite{falciano86} and E615~\cite{conway} for $\pi^-$ beam
interacting with tungsten ($W$) targets. The NA10 experiment used
three different beam energies: 140, 194 and 286 GeV, while E615
utilized a single beam energy of 252 GeV. Since the experiments were
done with tungsten targets, the cross sections per nucleon were
calculated by the weighted average of the $\pi^- p$ and $\pi^- n$
cross sections with 74 protons and 110 neutrons. Following the
experimental acceptance specified in Ref.~\cite{Lambertsen:2016wgj},
we apply the kinematic cuts listed in Table~\ref{tab:accpt}. The
results of NLO (red points) and NNLO (blue points) calculations
together with the measurements (black points) are shown in
Figs.~\ref{fig1_na10_140},~\ref{fig1_na10_194},~\ref{fig1_na10_286},
and ~\ref{fig1_e615}.

\begin{table}[tbp]   
\centering
\begin{tabular}{|c|c|c|c|}
\hline
\hline
 Experiment & Q (GeV) & $x_1$ & $x_F$ \\
\hline
\hline
NA10 & $4.05 \le Q \le 8.55$ & $0 \le x_1 \le 0.7$ & $0 \le x_F$ \\
\hline
E615 & $4.05 \le Q \le 8.55$ & $0.2 \le x_1 \le 1$ & $0 \le x_F$ \\
\hline
E866 & $4.5 \le Q \le 15$\textsuperscript{*} & $0 \le x_1 \le 0.7$ & $0 \le x_F$ \\
\hline
\hline
\multicolumn{4}{l}{\textsuperscript{*} \footnotesize{Excluding the
    $\Upsilon$ region $9 \le Q \le 10.7$ GeV.}}
\end{tabular}
\caption {Kinematic cuts applied for the experimental acceptance in
  the fixed-order pQCD calculation.}
\label{tab:accpt}
\end{table}

Overall, the calculated $\lambda$, $\mu$ and $\nu$ exhibit distinct
$q_T$ dependencies. At $q_T \rightarrow 0$, $\lambda$, $\mu$ and $\nu$
approach the values predicted by the collinear parton
model~\cite{drell}: $\lambda = 1$ and $\mu = \nu =0$. As $q_T$
increases, Figs.~\ref{fig1_na10_140}-\ref{fig1_e615} show that
$\lambda$ decreases toward its large-$q_T$ limit of $-1/3$ while $\nu$
increases toward $2/3$, for both $q\bar{q}$ and $qG$ processes shown
in Ref.~\cite{peng16}. The $q_T$ dependence of $\mu$ is relatively
mild compared to $\lambda$ and $\nu$. This is understood as a result
of some cancellation effect, to be discussed in
Sec.~\ref{sec:discussion}. Comparing the results of the NLO with the
NNLO calculation, $\lambda {\rm (NNLO)}$ is smaller than $\lambda
\rm{(NLO)}$ while $\mu$ and $\nu$ are very similar for NLO and
NNLO. The L-T violation, $1-\lambda-2\nu$, is zero in the NLO
calculation, and turns to be nonzero and positive in the NNLO
calculation.

As shown in Figs.~\ref{fig1_na10_140}-\ref{fig1_e615}, while some
general features of the NA10 and E615 data are described by the pQCD
calculations, there are notable differences between the data and
calculations. From the comparison between them, we find:

1) Perturbative QCD predicts that $\lambda$ drops as $q_T$ increases,
but the data do not show this trend. The expected upper bound of
$\lambda$, $\left| \lambda \right| \leq 1$, is sometimes exceeded by
the data~\cite{Lambertsen:2016wgj}. This could reflect the presence of
some systematic uncertainties in the data.

2) The agreement between the data and the pQCD calculation for the
$\mu$ parameter is quite reasonable for NA10, but less so for E615.

3) The increase of $\nu$ with $q_T$ observed in the NA10 data is in
good agreement with the pQCD calculation. However, the E615 data are
significantly higher than the calculation.

4) The amount of the L-T violation, $1-\lambda-2\nu$, for the data is
much larger than the prediction from the NNLO pQCD. Moreover, the sign
of this violation is negative for the data, but positive for the
pQCD. This apparent discrepancy could be partly caused by the
unphysical values of $\lambda$ from the data, as $\lambda$ should not
exceed 1.

Regarding these findings two remarks are in order. First, pQCD
predicts a sizable magnitude for $\nu$, comparable to the
data. Therefore, in order to extract the value of the nonperturbative
Boer-Mulders function from the measured data of
$\nu$~\cite{boer99,Zhang:2008nu,Barone:2009hw}, contributions from the
pQCD effect must be taken into account. Second, the pQCD calculation
for $\mu$ tends to overestimate the NA10 data but underestimate the
E615 data. As we will see in Sec.~\ref{sec:results_newexp}, $\mu$ has
a strong dependence on $x_F$. The incomplete information on the $x_F$
acceptance of the experiments needed for the calculation could
contribute to the discrepancy.

\begin{figure}[htbp]
\includegraphics[width=1.0\columnwidth]{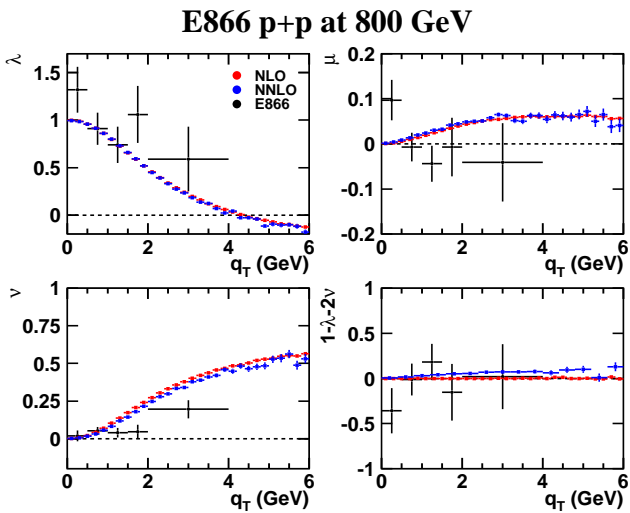}
\caption{Comparison of NLO (red points) and NNLO (blue points)
  fixed-order pQCD calculations with the E866 $p+p$ D-Y data at 800
  GeV~\cite{zhu} (black points) for $\lambda$, $\mu$, $\nu$ and
  $1-\lambda-2\nu$.}
\label{fig1_e866p}
\end{figure}

\begin{figure}[htbp]
\includegraphics[width=1.0\columnwidth]{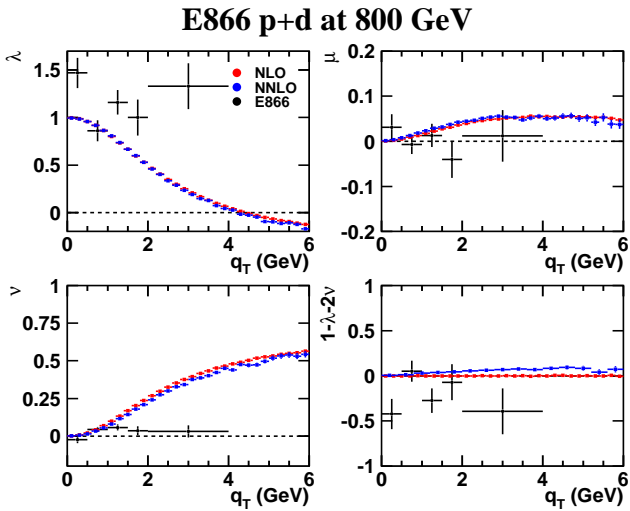}
\caption{Same as Fig.~\ref{fig1_e866p}, but for E866 data~\cite{zhu}
  with a liquid deuterium target.}
\label{fig1_e866d}
\end{figure}

The $q_T$ dependencies of the angular distribution parameters of
800-GeV $p+p$ and $p+d$ D-Y are calculated and compared with the E866
measurements~\cite{zhu} in Figs.~\ref{fig1_e866p} and
~\ref{fig1_e866d}. Given the large experimental uncertainty, the $p+p$
data in Fig.~\ref{fig1_e866p} are not in disagreement with the
calculation. For Fig.~\ref{fig1_e866d}, where the $p+d$ data have
smaller uncertainties, the agreement between data and the calculation
is rather poor. In particular, the data on $\lambda$ are in general
larger than 1, violating the expected upper bound for
$\lambda$~\cite{Lambertsen:2016wgj}.


That the $\nu$ data are less than the pQCD prediction in
Figs.~\ref{fig1_e866p} and ~\ref{fig1_e866d}, suggests a negative
contribution from the Boer-Mulders effect in the proton-induced
DY. This is opposite to the situation in Fig.~\ref{fig1_e615}, where
the $\nu$ data are more positive than the pQCD, suggesting a positive
contribution from the Boer-Mulders function in the pion-induced
DY. Since the contribution of the Boer-Mulders effect in $\nu$ is
proportional to the product of the individual Boer-Mulders functions
of quarks and anti-quarks in the colliding hadrons, the proton D-Y
data would imply that the sea-quark Boer-Mulders function has a sign
opposite to that of the valence Boer-Mulders function in the
proton~\cite{transversity2014}. The pion data from
Fig.~\ref{fig1_e615} suggests that the pion valence Boer-Mulders
function has the same sign as the proton valence Boer-Mulders
function~\cite{transversity2014}.

The NNLO calculations predict a positive $1-\lambda-2\nu$ at NNLO while
the data are consistent with zero for the proton target and slightly
negative for the deuteron one. The negative values of $1-\lambda-2\nu$
for $p+d$ data are similar to the case for the pion D-Y data shown in
Figs.~\ref{fig1_na10_140}-\ref{fig1_e615}. In
Sec.~\ref{sec:discussion}, we will discuss why $1-\lambda-2\nu$ must
be positive from the perspective of a geometric approach.

\section{pQCD calculations for the COMPASS and SeaQuest experiments}
\label{sec:results_newexp}

There are two ongoing fixed-target D-Y experiments which have
collected new data on the lepton angular distributions. The first one
is the COMPASS experiment at CERN~\cite{COMPASSII}, running with
190-GeV $\pi^-$ beam and transversely-polarized $\rm{NH}_3$ target and
unpolarized aluminum ($Al$) and tungsten ($W$) nuclear targets. The
transverse-momentum-dependent Sivers asymmetry in the polarized D-Y
process was reported recently~\cite{compass}, and high-statistics
unpolarized D-Y data on the $W$ target have also been collected. The
second one is the SeaQuest experiment at Fermilab~\cite{E906}, aiming
at the measurement of $\bar{d}(x)/\bar{u}(x)$ ratio at
intermediate-$x$ region via the D-Y process. It has taken data with
the 120-GeV proton beam on unpolarized hydrogen, deuterium and various
nuclear targets. Both COMPASS and SeaQuest experiments have collected
data on the lepton angular distributions of the D-Y process. The final
results are expected to be available in the near future. In addition,
the extension of the SeaQuest experiment, the E1039
experiment~\cite{e1039}, expects to take more data relevant to the
angular distributions in the near future.

Here we present the results of the angular coefficients $\lambda$,
$\mu$ and $\nu$ as a function of $q_T$ in various bins of $Q$ and
$x_F$. There are three bins for $Q$ in the range of 4.0--7.0 GeV, as
well as three bins for $x_F$ in the range of 0--0.6. These results
could be convoluted by the COMPASS and SeaQuest spectrometer
acceptances later for a direct comparison with experimental
data. Since there are no significant difference between the NLO and
NNLO results, we present only the results from the NLO calculation to
illustrate the major features.

The mean values of $Q$ and $x_F$ in each bin are listed in
Tables~\ref{tab:mean_Q} and ~\ref{tab:mean_xF}. The pQCD calculations
show that the $q \bar{q}$ process dominates over the whole $q_T$
region for the $\pi^-$-induced COMPASS experiment while the $qG$
process becomes more important for $q_T>1$ GeV in the proton-induced
SeaQuest experiment. Through this study, the $Q$- and
$x_F$-dependencies of $\lambda$, $\mu$ and $\nu$ are also
investigated.


\begin{table}[tbp]   
\caption {Mean values of $Q$ and $x_F$ in each $Q$ bin calculated
  for COMPASS and SeaQuest.}
\label{tab:mean_Q}
\begin{center}
\begin{tabular}{|c|c|c|c|c|}
\hline
\hline
Bin & \multicolumn{2}{c|} {COMPASS} &  \multicolumn{2}{c|} {SeaQuest} \\
\hline
\hline
  & $\langle Q \rangle$ (GeV) & $\langle x_F \rangle $ & $\langle Q \rangle$ (GeV) & $\langle x_F \rangle $ \\
\hline
$Q=4-5$ GeV & 4.42 & 0.32 & 4.36 & 0.24 \\
\hline
$Q=5-6$ GeV & 5.43 & 0.32 & 5.36 & 0.23 \\
\hline
$Q=6-7$ GeV & 6.43 & 0.32 & 6.36 & 0.22 \\
\hline
\hline
\end{tabular}
\end{center}
\end{table}

\begin{table}[tbp]   
\caption {Mean values of $Q$ and $x_F$ in each $x_F$ bin calculated
  for COMPASS and SeaQuest.}
\label{tab:mean_xF}
\begin{center}
\begin{tabular}{|c|c|c|c|c|}
\hline
\hline
Bin & \multicolumn{2}{c|} {COMPASS} &  \multicolumn{2}{c|} {SeaQuest} \\
\hline
\hline
  & $\langle Q \rangle$ (GeV) & $\langle x_F \rangle $ & $\langle Q \rangle$ (GeV) & $\langle x_F \rangle $ \\
\hline
$x_F=0.0-0.2$ & 5.01 & 0.10 & 4.56 & 0.10 \\
\hline
$x_F=0.2-0.4$ & 5.06 & 0.30 & 4.55 & 0.29 \\
\hline
$x_F=0.4-0.6$ & 5.10 & 0.49 & 4.54 & 0.48 \\
\hline
\hline
\end{tabular}
\end{center}
\end{table}

\begin{figure}[htbp]
\centering
\subfloat[]
{\includegraphics[width=1.0\columnwidth]{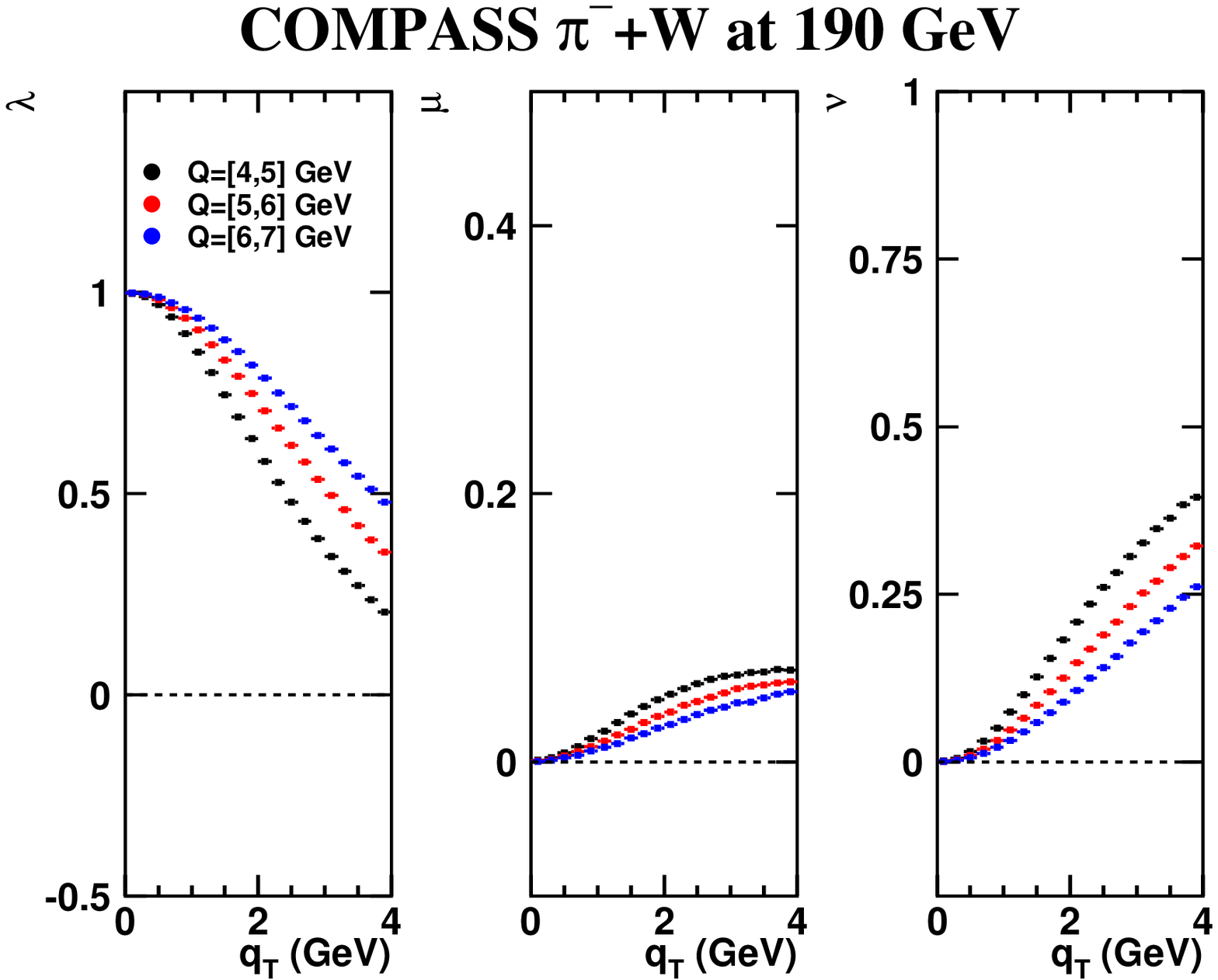}}\\
\subfloat[]
{\includegraphics[width=1.0\columnwidth]{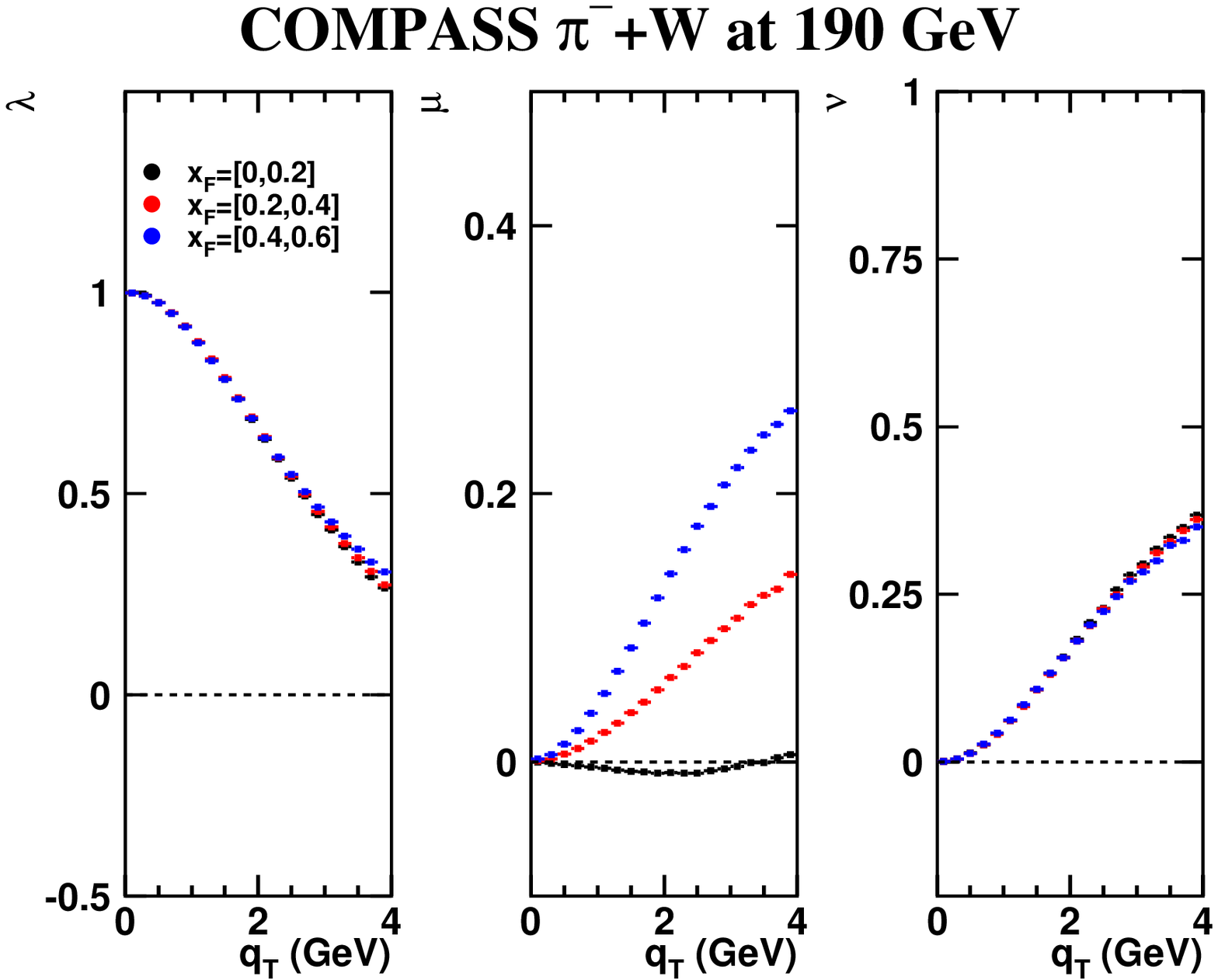}}
\caption
[\protect{}] {(a) NLO pQCD results of $\lambda$, $\mu$, and $\nu$ as a
  function of $q_T$ at several $Q$ bins and $x_F>0$ for D-Y production
  off the tungsten target with 190-GeV $\pi^-$ beam in the COMPASS
  experiment. (b) Same as (a) but at several $x_F$ bins and $4<Q<9$
  GeV.}
\label{fig2_compass}
\end{figure}
\begin{figure}[htbp]
\centering
\subfloat[]
{\includegraphics[width=1.0\columnwidth]{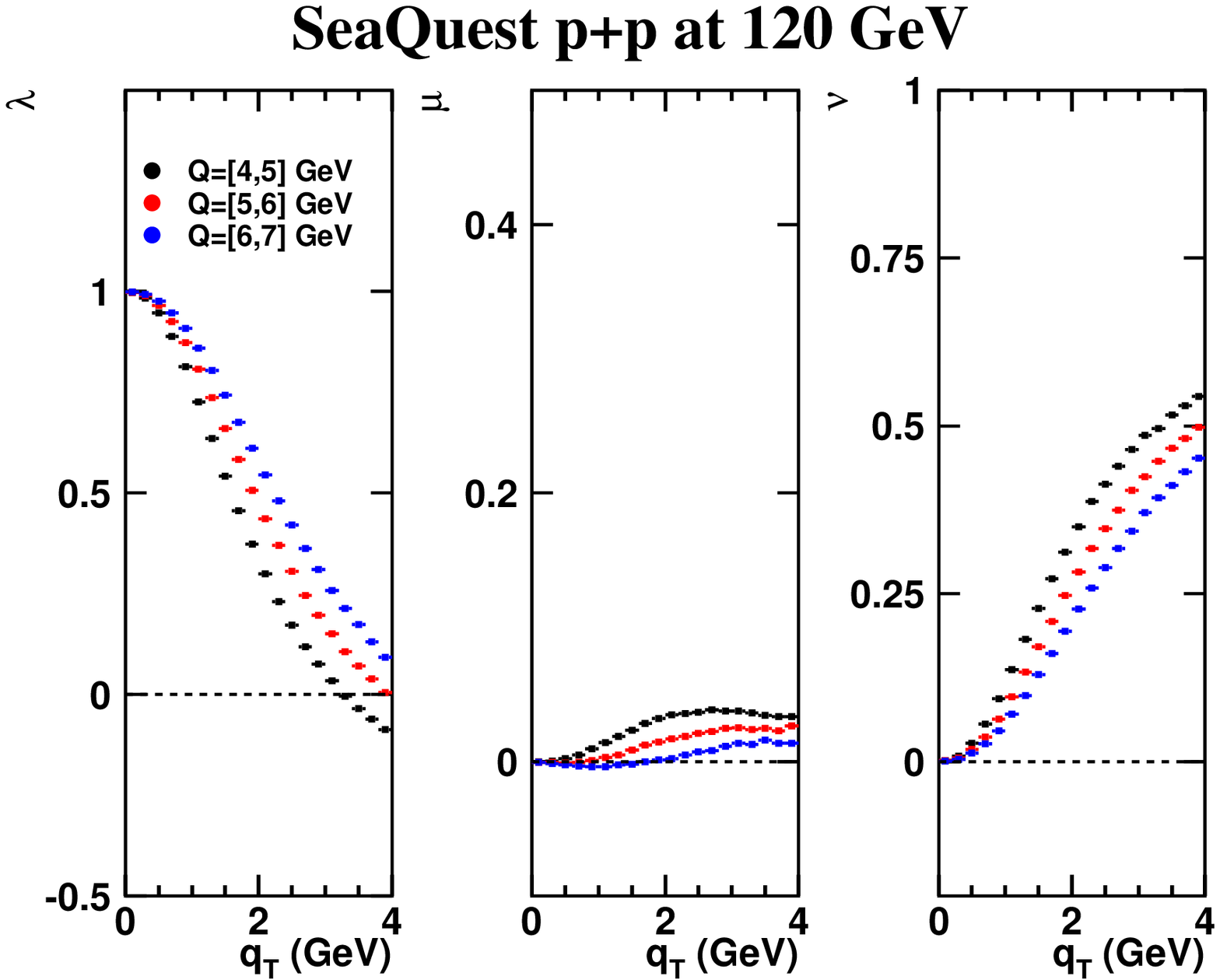}}\\
\subfloat[]
{\includegraphics[width=1.0\columnwidth]{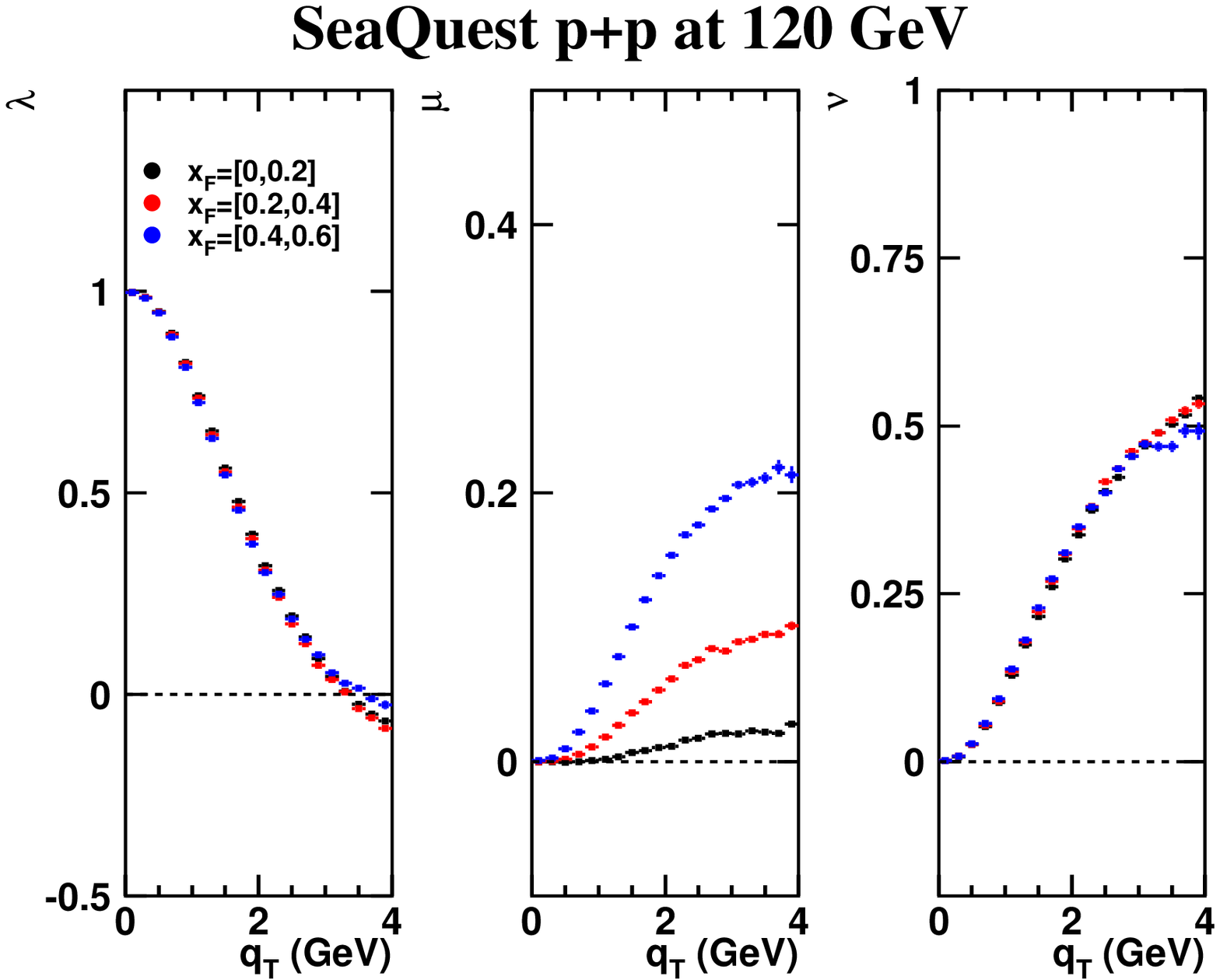}}
\caption
[\protect{}] {(a) NLO pQCD results of $\lambda$, $\mu$, and $\nu$ as a
  function of $q_T$ at several $Q$ bins and $x_F>0$ for D-Y production
  off the proton target with 120-GeV proton beam in the SeaQuest
  experiment. (b) Same as (a) but at several $x_F$ bins and $4<Q<9$
  GeV.}
\label{fig2_e906}
\end{figure}

Figures~\ref{fig2_compass} and ~\ref{fig2_e906} show $\lambda$, $\mu$
and $\nu$ as a function of $q_T$ for various bins of $Q$ and
$x_F$. The $q_T$ distributions of $\lambda$ and $\nu$ parameters
depend sensitively on $Q$, but only weakly on $x_F$. As for $\mu$, its
$q_T$ distribution has strong dependencies on $x_F$ and on $Q$. In
particular, the magnitude of $\mu$ is small when $x_F$ is close to 0
and its sign could even turn negative at some $q_T$ region. As $x_F$
increases, the magnitude of $\mu$ increases pronouncedly.

In perturbative QCD at $\mathcal{O}(\alpha_S)$, ignoring the intrinsic
transverse momenta of the colliding partons, the $\lambda$ and $\nu$
coefficients in the Collins-Soper frame for the $q \bar q \to \gamma^* G$ annihilation
process~\cite{collins,boer06,berger} and the $qG \to \gamma^* q$
Compton process~\cite{falciano86,thews,lindfors} are given as
\begin{align}
\lambda &=  \frac{2 Q^2-q_T^2}{2Q^2+ 3q_T^2} & \nu &= 
\frac{2 q_T^2}{2Q^2+ 3q_T^2} & &(q\bar q)
\nonumber \\
\lambda &= \frac{2Q^2-5q_T^2}{2Q^2+15q_T^2} & \nu &= 
\frac{10q_T^2}{2Q^2+15q_T^2} & &(qG),
\label{eq:eq11}
\end{align}
where $q_T$ and $Q$ are the transverse momentum and mass,
respectively, of the dilepton. While the expression for $q \bar q \to
\gamma^* G$ is exact, that for $qG \to \gamma^* q$ is obtained with
some approximation. Equation (\ref{eq:eq11}) shows that $\lambda$ and
$\nu$ scale with the dimensionless $q_T/Q$ in these pQCD NLO
expressions. Nevertheless there is no $q_T/Q$ scaling for the $\mu$
parameter in NLO pQCD.

\begin{figure}[htbp]
\centering
\subfloat[]
{\includegraphics[width=1.0\columnwidth]{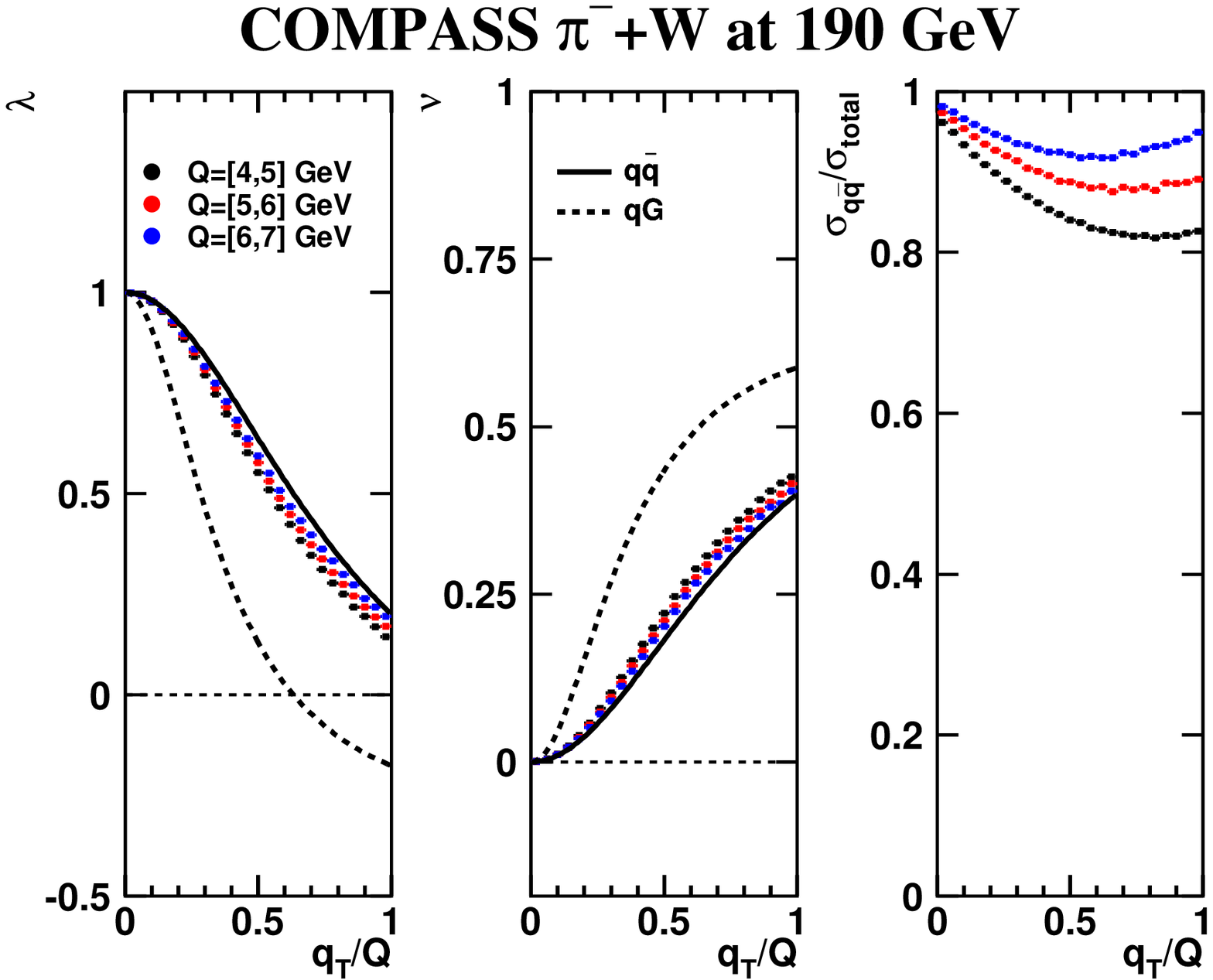}}\\
\subfloat[]
{\includegraphics[width=1.0\columnwidth]{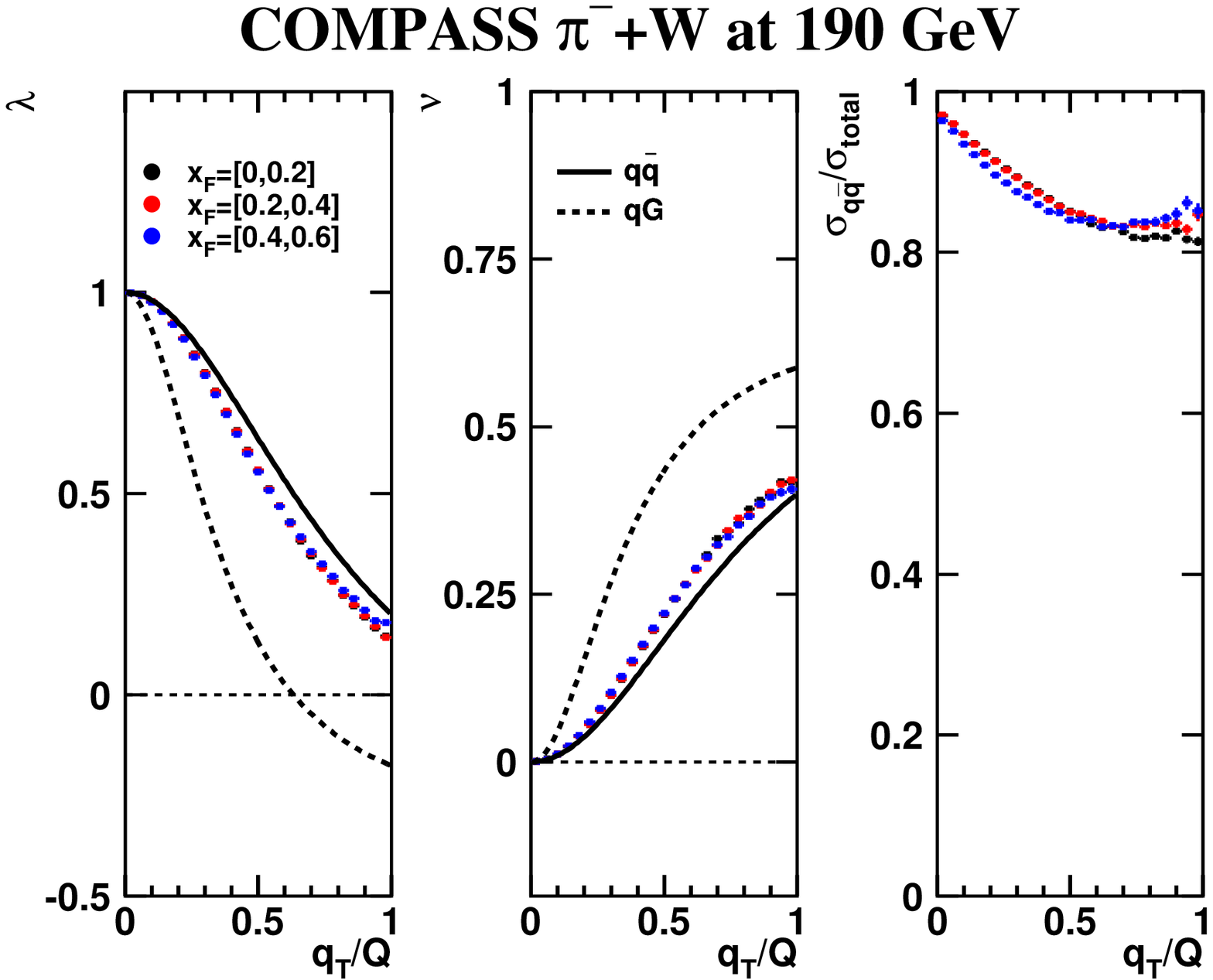}}
\caption
[\protect{}] {(a) NLO pQCD results of $\lambda$, $\nu$ and the
  fractions of $q \bar{q}$-process contribution in the total cross
  sections as a function of scaled transverse momentum $q_T/Q$ for D-Y
  production off the nuclear tungsten target with 190-GeV $\pi^-$ beam
  in the COMPASS experiment. The NLO pQCD expressions of $q \bar{q}$
  and $qG$ processes are denoted by the solid and dashed lines
  respectively. (b) Same as (a) but at several $x_F$ bins and $4<Q<9$
  GeV.}
\label{fig3_compass}
\end{figure}

\begin{figure}[htbp]
\centering
\subfloat[]
{\includegraphics[width=1.0\columnwidth]{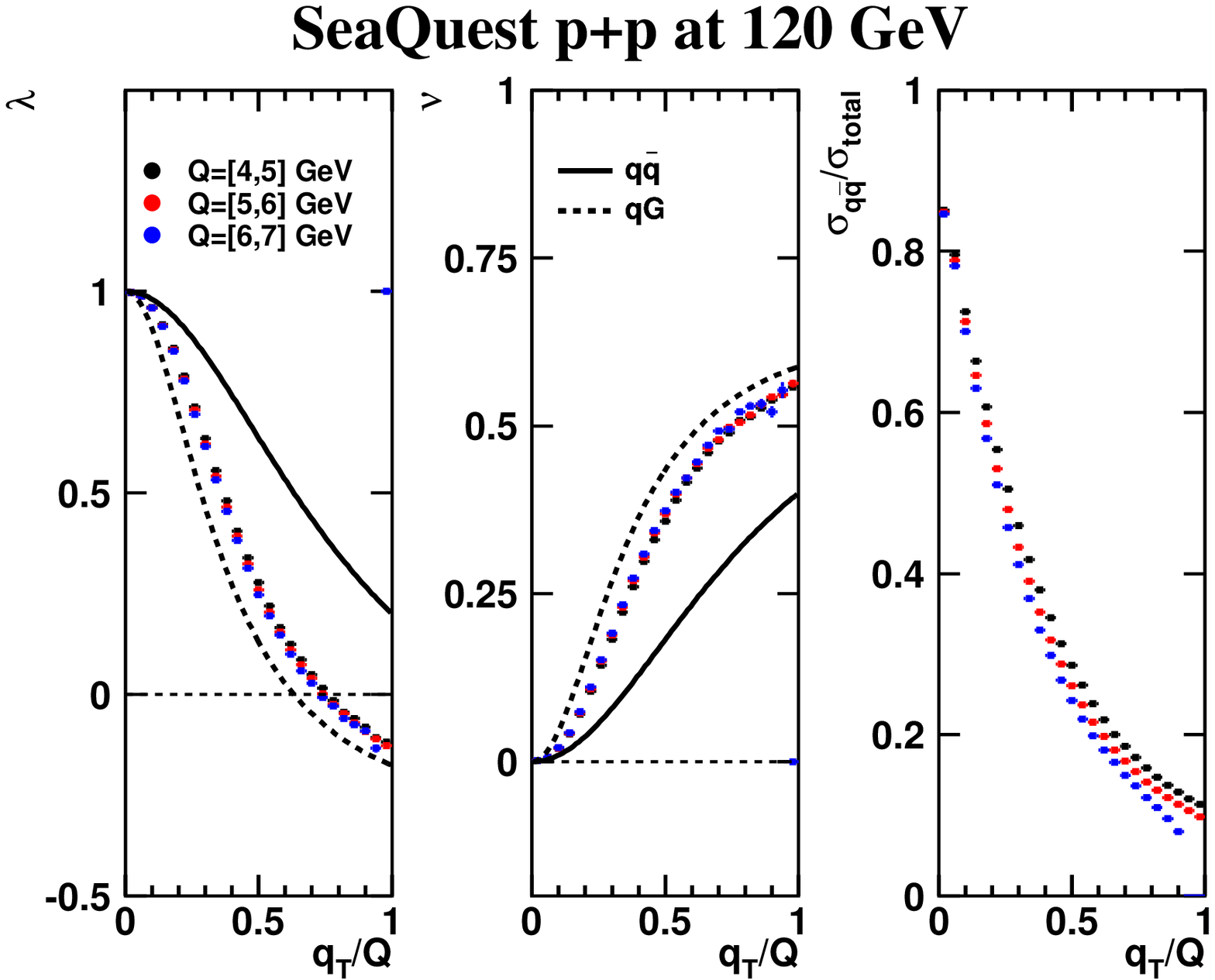}}\\
\subfloat[]
{\includegraphics[width=1.0\columnwidth]{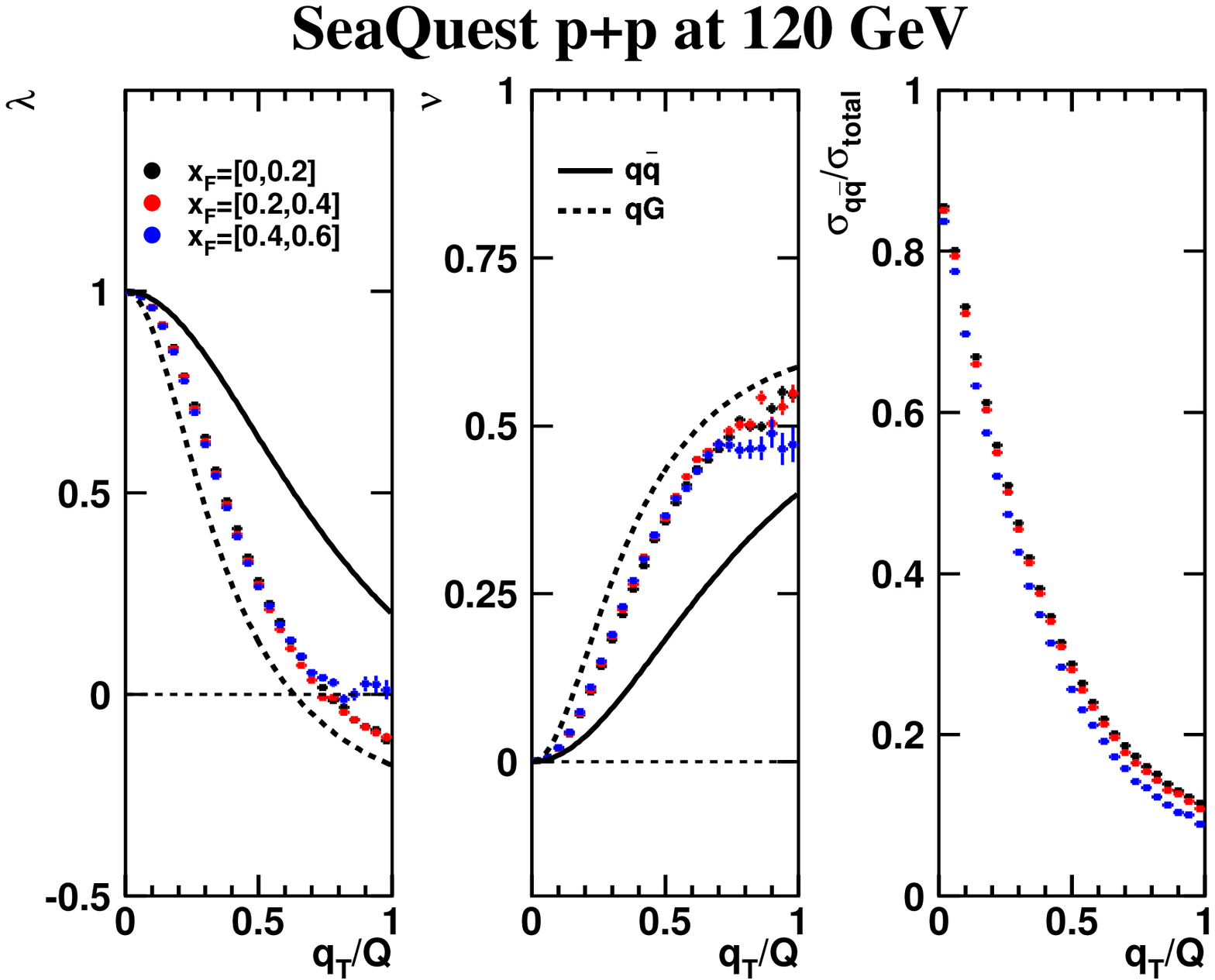}}
\caption
[\protect{}]{(a) NLO pQCD results of $\lambda$, $\nu$ and the
  fractions of $q \bar{q}$-process contribution in the total cross
  sections as a function of scaled transverse momentum $q_T/Q$ for D-Y
  production off the proton target with 120-GeV proton beam in the
  SeaQuest experiment. The NLO pQCD expressions of $q \bar{q}$ and
  $qG$ processes are denoted by the solid and dashed lines
  respectively. (b) Same as (a) but at several $x_F$ bins and $4<Q<9$
  GeV. It is noted that the rough structure at large $q_T/Q$ region of
  the results for $x_F=$0.4 -- 0.6 (blue points) is likely due to the
  fluctuation of calculations with $Q>7$ GeV near the edge of the
  phase space. The structure is expected to be removed, if one
  requires $Q<7$ GeV as the top figure.}
\label{fig3_e906}
\end{figure}

Figures~\ref{fig3_compass} and~\ref{fig3_e906} show the NLO
calculations of $\lambda$ and $\nu$ for COMPASS and SeaQuest as a
function of the variable $q_T/Q$ in the various $Q$ and $x_F$
bins. The corresponding expressions for the $q \bar{q}$ and $q G$
processes in Eq.~(\ref{eq:eq11}) are denoted by the solid and dashed
lines. Comparing Figs.~\ref{fig3_compass} and~\ref{fig3_e906} with
Figs.~\ref{fig2_compass} and~\ref{fig2_e906}, the $\lambda$ and $\nu$
values for different $Q$ bins now converge into a common curve when
they are plotted as a function of $q_T/Q$. This is consistent with the
$q_T/Q$ scaling behavior of Eq.~(\ref{eq:eq11}).

Figures~\ref{fig3_compass} and~\ref{fig3_e906} also display the
fractions of the NLO cross sections due to the $q \bar{q}$ process for
COMPASS and SeaQuest. The dominance of the $q \bar{q}$ process in the
$\pi^-$-induced D-Y at COMPASS explains why the pQCD results for
$\lambda$ and $\nu$ are very close to the solid $q \bar{q}$ lines. In
contrast, the proton-induced D-Y in SeaQuest has large contributions
from the $q G$ process, resulting in the $\lambda$ and $\nu$ closer to
the dashed $q G$ lines.

\begin{figure}[htbp]
\centering
\subfloat[]
{\includegraphics[width=1.0\columnwidth]{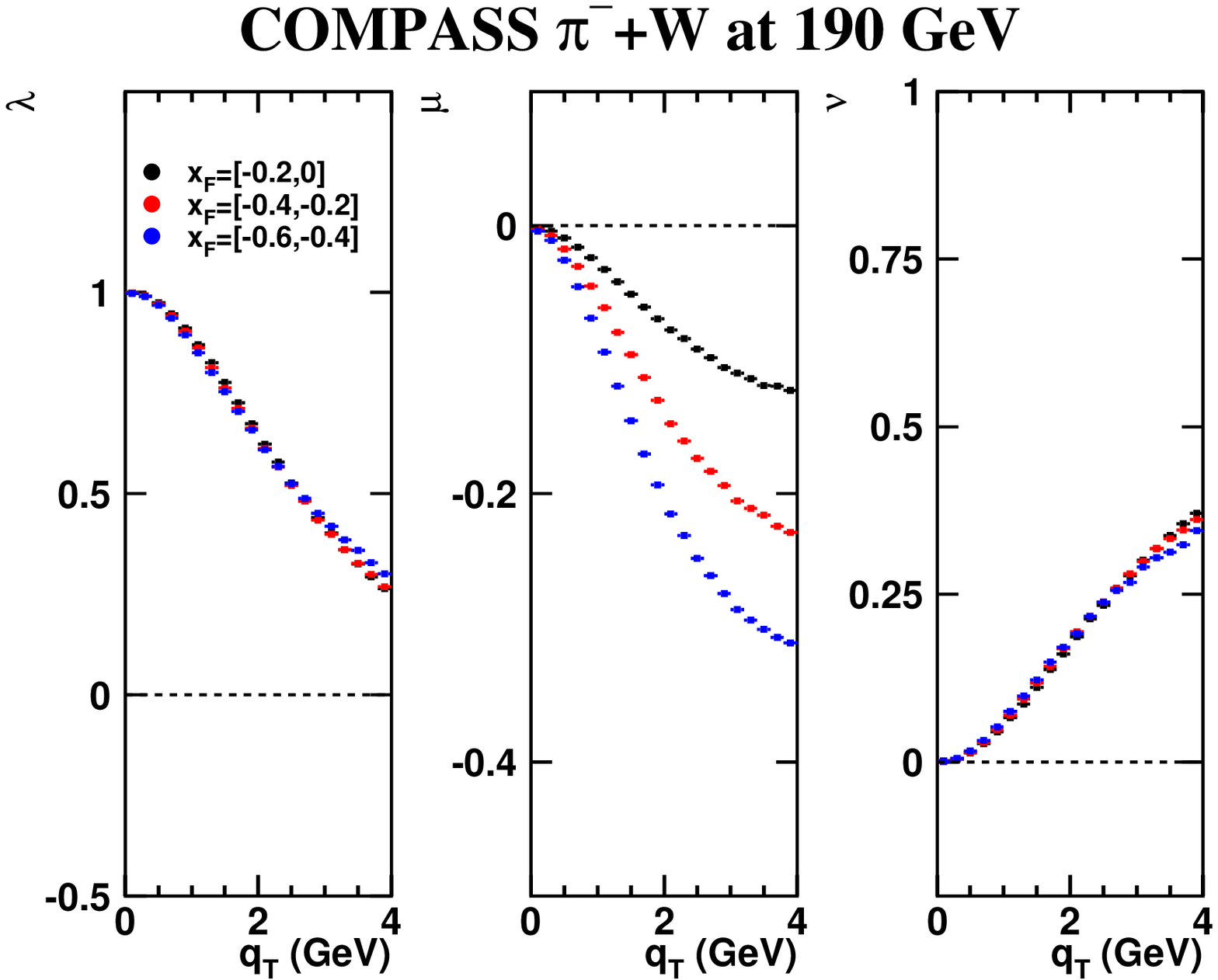}}\\
\subfloat[]
{\includegraphics[width=1.0\columnwidth]{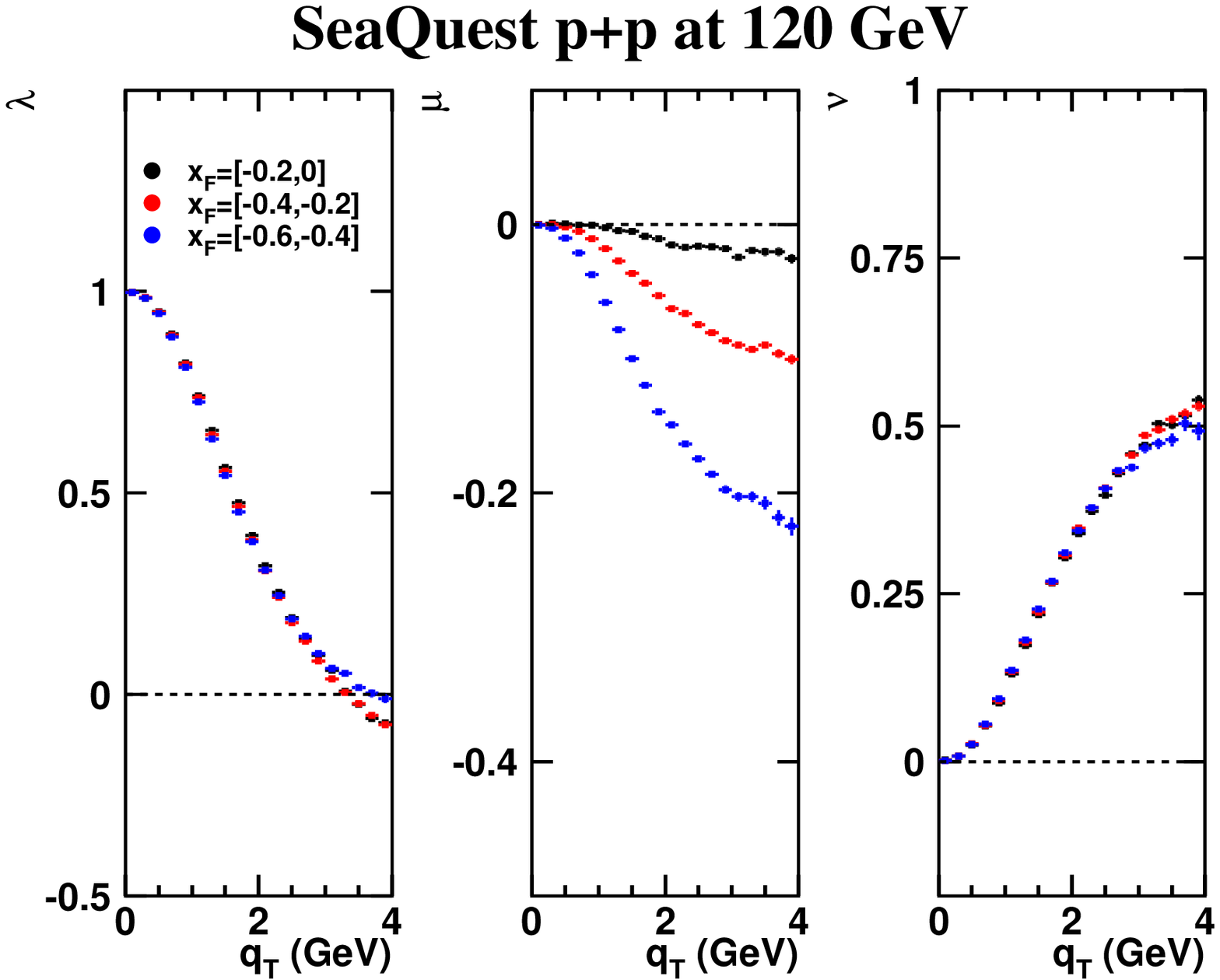}}
\caption
[\protect{}] {(a) NLO pQCD results of $\lambda$, $\mu$ and $\nu$ as a
  function of transverse momentum $q_T$ at several negative $x_F$ bins
  and $4<Q<9$ GeV for D-Y production off the nuclear tungsten target
  with 190-GeV proton beam in COMPASS experiment. (b) Same results of
  (a) for D-Y production off the proton target with 120-GeV proton beam
  in SeaQuest experiment.}
\label{fig2b}
\end{figure}

In comparison, we plot the $q_T$ distributions of $\lambda$, $\mu$ and
$\nu$ in the negative $x_F$ (-0.6 -- 0) for COMPASS and SeaQuest in
Fig.~\ref{fig2b}. The $\lambda$ and $\nu$ remain the same as that in
$x_F>0$ while $\mu$ turns mostly negative.

\section{Geometric model}
\label{sec:discussion}

As seen above, the existing D-Y data of lepton angular distributions
can be reasonably well described by the NLO and NNLO pQCD
calculations. Various salient features of $Q$ and $x_F$ dependencies
as well as $q_T/Q$ scaling are observed in the predicted results of
$\lambda$, $\mu$ and $\nu$ parameters for COMPASS and SeaQuest
experiments based on NLO pQCD. It is of interest to check if these
features could be understood using the geometric approach developed in
Refs.~\cite{peng16,chang17}.

\begin{figure}[htb]
\includegraphics[width=0.8\columnwidth]{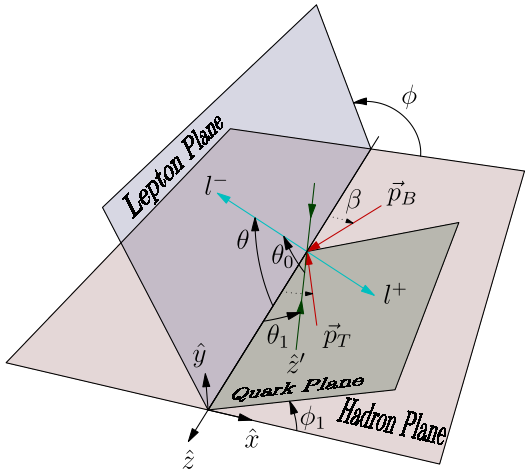}
\caption
[\protect{}]{Definition of the Collins-Soper frame and various angles
  and planes in the rest frame of $\gamma^*$. The hadron plane is
  formed by $\vec P_B$ and $\vec P_T$, the momentum vectors of the
  beam (B) and target (T) hadrons. The $\hat x$ and $\hat z$ axes of
  the Collins-Soper frame both lie in the hadron plane with the $\hat
  z$ axis bisecting the $\vec P_B$ and $- \vec P_T$ vectors.  The
  quark ($q$) and antiquark ($\bar q$) annihilate collinearly with
  equal momenta to form $\gamma^*$, while the quark momentum vector
  $\hat z^\prime$ and the $\hat z$ axis form the quark plane. The
  polar and azimuthal angles of $\hat z^\prime$ in the Collins-Soper
  frame are $\theta_1$ and $\phi_1$. The $l^-$ and $l^+$ are emitted
  back-to-back with $\theta$ and $\phi$ as the polar and azimuthal
  angles for $l^-$.}
\label{fig_geom}
\end{figure}

Here we sketch the geometric approach of
Refs.~\cite{peng16,chang17}. As illustrated in Fig.~\ref{fig_geom}, we
define three different planes, the hadron plane, the quark plane, and
the lepton plane, in the Collins-Soper frame. In the $\gamma^*$ rest
frame, the beam and target hadron momenta, $\vec P_B$ and $\vec P_T$
form the ``hadron plane'' on which the $\hat z$ axis, bisecting the
$\vec P_B$ and $- \vec P_T$ vectors, lies. A pair of collinear $q$ and
$\bar q$ with equal momenta annihilate into a $\gamma^*$. The momentum
unit vector of $q$ is defined as $\hat z^\prime$, and the ``quark
plane" is formed by the $\hat z^\prime$ and $\hat z$ axes. Finally,
the ``lepton plane'' is formed by the momentum vector of $l^-$ and the
$\hat z$ axis. The polar and azimuthal angles of the $\hat z^\prime$
axis in the Collins-Soper frame are denoted as $\theta_1$ and
$\phi_1$. As shown in Refs.~\cite{peng16,chang17}, the angular
coefficients $A_i$ in Eq.~(\ref{eq:eq3}) can be expressed in term of
$\theta_1$ and $\phi_1$ as follows:
\begin{eqnarray}
A_0 &=&  \langle\sin^2\theta_1\rangle \nonumber \\
A_1 &=& \frac{1}{2} \langle\sin 2\theta_1\cos \phi_1\rangle \nonumber \\
A_2 &=&  \langle\sin^2\theta_1 \cos 2\phi_1\rangle.
\label{eq:eq8}
\end{eqnarray}
The $\langle \cdot \cdot \cdot \rangle$ in Eq.~(\ref{eq:eq8}) is a
reminder that the measured values of $A_i$ at a given kinematic bin
are averaged over events having particular values of $\theta_1$ and
$\phi_1$.

As discussed in Refs.~\cite{peng16,chang17}, up to NLO
($\mathcal{O}(\alpha_S)$) in pQCD, the quark plane coincides with the
hadron plane and $\phi_1=0$. Therefore $A_0=A_2$ or
$1-\lambda-2\nu=0$, i.e., the L-T relation is satisfied. Higher order
pQCD processes allow the quark plane to deviate from the hadron plane,
i.e., $\phi_1 \neq 0$, leading to the violation of the L-T
relation. For a nonzero $\phi_1$, Eq.~(\ref{eq:eq8}) shows that $A_2
< A_0$. Therefore, when the L-T relation is violated, $A_0$ must be
greater than $A_2$ or, equivalently, $1 - \lambda - 2\nu >0$. This
expectation of $1 - \lambda - 2\nu >0$ in the geometric approach is in
agreement with the results of NNLO pQCD calculations shown in
Figs.~\ref{fig1_na10_140}-\ref{fig1_e866d}. The geometric approach
offers a simple interpretation for this result.

\begin{figure}[tb]
\centering
\subfloat[]
{\includegraphics[width=0.23\textwidth]{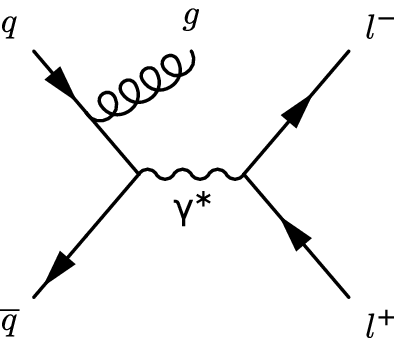}}
\subfloat[]
{\includegraphics[width=0.23\textwidth]{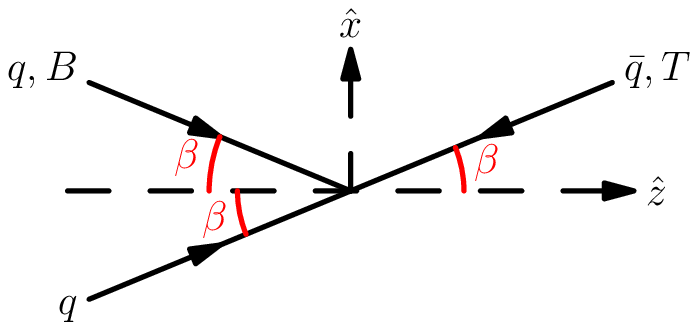}}
\qquad
\subfloat[]
{\includegraphics[width=0.23\textwidth]{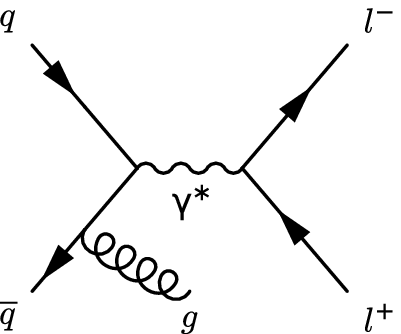}}
\subfloat[]
{\includegraphics[width=0.23\textwidth]{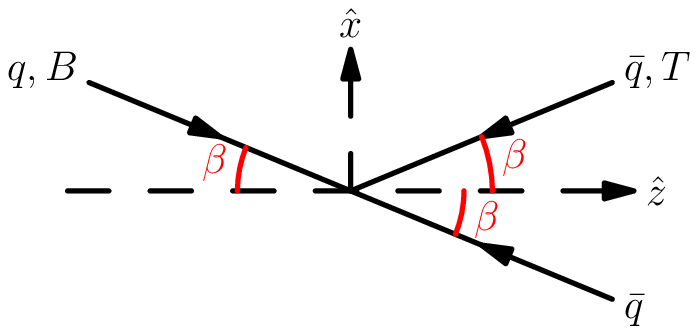}}
\caption
[\protect{}]{(a) Feynman diagram for $q - \bar q$ annihilation where a
  gluon is emitted from a quark in the beam hadron. (b) Momentum
  vectors for $q$ and $\bar q$ in the C-S frame before and after gluon
  emission. The momentum direction of $q$ is now collinear with that
  of $\bar q$. (c) Feynman diagram for the case where a gluon is
  emitted from an antiquark in the target hadron. (d) Momenta vectors
  for $q$ and $\bar q$ in the C-S frame before and after gluon
  emission for diagram (c).}
\label{fig6}
\end{figure}

Figures ~\ref{fig2_compass}(b) and ~\ref{fig2_e906}(b) show that the
$q_T$ dependencies for $\lambda$ and $\nu$ are insensitive to the
value of $x_F$. In contrast, the $\mu$ parameter depends sensitively
on $x_F$. This striking difference between the $\lambda$, $\mu$ and
$\nu$ parameters can be understood in the geometric approach. At the
next-to-leading order (NLO), $\mathcal{O}(\alpha_S)$, a hard gluon or
a quark (antiquark) is emitted so that $\gamma^*$ acquires nonzero
$q_T$. Figure~\ref{fig6}(a) shows a diagram for the $q - \bar q$
annihilation process in which a gluon is emitted from the quark in the
beam hadron. In this case, the momentum vector of the quark is
modified such that it becomes opposite to the antiquark's momentum
vector in the rest frame of $\gamma^*$ [Fig.~\ref{fig6}(b)]. Since the
antiquark's momentum is the same as the target hadron's, the
$\hat z^\prime$ axis is along the direction of $- \vec p_T$.  From
Fig.~\ref{fig_geom}, it is evident that $\theta_1 = \beta$ and $\phi_1
= 0$ in this case. An analogous diagram in which the gluon is emitted
from the antiquark in the target hadron is shown in
Fig.~\ref{fig6}(c). In this case, $\theta_1 = \beta$ while $\phi_1 =
\pi$. Table~\ref{tab:angles} lists the values of $\theta_1$ and
$\phi_1$ for four cases of different combination of hadron and quark
types from which the gluon is emitted~\cite{chang17}.

\begin{table}[tbp]  
\caption {Angles $\theta_1$ and $\phi_1$ for four cases of gluon emission
in the $q - \bar q$ annihilation process at order-$\alpha_s$. The signs of
$A_0$, $A_1 (\mu)$, $A_2 (\nu)$ for the four cases are also listed.}
\label{tab:angles}
\begin{center}
\begin{tabular}{|c|c|c|c|c|c|c|}
\hline
\hline
case & gluon emitted from & $\theta_1$ & $\phi_1$ & $A_0$ & $A_1 (\mu)$ & $A_2 (\nu)$ \\
\hline
\hline
1 & beam quark & $\beta$ & 0 & + & + & + \\
\hline
2 & target antiquark & $\beta$ & $\pi$ & + & $-$ & + \\
\hline
3 & beam antiquark & $\pi - \beta$ & 0 & + & $-$ & + \\
\hline
4 & target quark & $\pi - \beta$ & $\pi$ & + & + & + \\
\hline
\hline
\end{tabular}
\end{center}
\end{table}

Table~\ref{tab:angles} shows that the sign of $\mu$ could be either
positive or negative, depending on which parton and hadron the gluon
is emitted from. Hence, one expects some cancellation effects for
$\mu$ among contributions from various processes. Each process is
weighted by the corresponding density distributions for the
interacting partons. At $x_F \sim 0$, the momentum fraction carried by
the beam parton ($x_B$) is comparable to that of the target parton
($x_T$). Therefore, the weighting factors for various processes are of
similar magnitude and the cancellation effect could be very
significant, resulting in a small value of $\mu$. On the other hand,
as $x_F$ increases toward 1, $x_B$ becomes much larger than $x_T$. In
this case the weighting factors are now dominated by fewer processes,
resulting in less cancellation and a larger value of $\mu$. This
explains why the $\mu$ parameter exhibits a strong $x_F$ dependence in
Figs.~\ref{fig2_compass}(b),~\ref{fig2_e906}(b) and~\ref{fig2b}.

Table~\ref{tab:angles} also shows that $A_0$ and $A_2$ have the same
sign (positive) for all four cases. This implies the absence of
$x_F$-dependent cancellation effect for them. Hence $\lambda$ and
$\nu$ have very weak $x_F$ dependencies, as shown in
Figs.~\ref{fig2_compass}(b),~\ref{fig2_e906}(b)
and~\ref{fig2b}. Therefore, the observed strong rapidity dependence
for $\mu$ and weak rapidity dependence for $\lambda$ and $\nu$ in pQCD
calculation can be nicely described by the geometric picture. In
addition, considering the strong $x_F$-dependence for the $q_T$
distribution of $\mu$ parameters, it will be instructive for the
experiments to measure the $q_T$ dependence of $\mu$ at several $x_F$
regions, instead of integrating over the entire $x_F$.

The NLO pQCD expressions of $\lambda$ and $\nu$ as a function of $q_T$
in Eq.~(\ref{eq:eq11}) have been derived based on a geometric picture
of collision geometry in the parton
level~\cite{peng16,chang17}. Within the geometric picture, the $A_0$
and $A_2$ at NLO are equal to $\langle\sin^2\theta_1\rangle$
(Eq.~(\ref{eq:eq8})) with $\phi_1=0$.  Given $q_T/Q=\tan \theta_1$ or
$-\tan\theta_1$, the scaling of $A_0$ and $A_2$ (equivalently
$\lambda$ and $\nu$) with $q_T/Q$ could also be understood.

\begin{figure}[htbp]
\includegraphics[width=1.0\columnwidth]{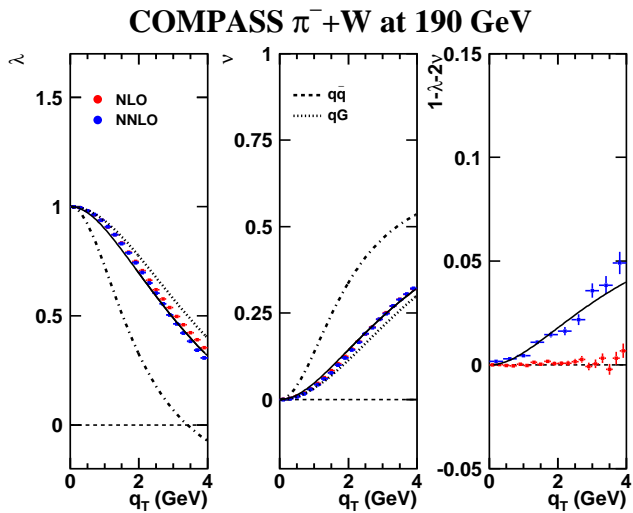}
\caption{NLO (red points) and NNLO (blue points) pQCD results of
  $\lambda$, $\mu$, $\nu$ and $1-\lambda-2\nu$ as a function of $q_T$
  at the kinematic bin of $5<Q<6$ GeV and $0.2<x_F<0.4$ for D-Y
  production off the tungsten target with 190-GeV $\pi^-$ beam in
  COMPASS experiment. The NLO pQCD expressions of $q \bar{q}$ and $qG$
  processes are denoted by the dotted and dash-dotted lines
  respectively. The solid curves correspond to the fit results
  described in the text.}
\label{fig4_compass}
\end{figure}

Figure~\ref{fig4_compass} shows both NLO (red points) and NNLO (blue
points) pQCD results of $\lambda$, $\nu$ and $1-\lambda-2\nu$ as a
function of $q_T$ at the kinematic bin of $5<Q<6$ GeV and
$0.2<x_F<0.4$ for the COMPASS experiment. The corresponding NLO pQCD
expressions of $q_T$ dependence for $q\bar{q}$ and $qG$ subprocesses
in Eq.~(\ref{eq:eq11}) are drawn as dotted and dotted-dash
curves. Assuming the fraction of these two processes is $q_T$
independent, a best-fit to the NNLO results of $\lambda$ yields the
fraction of $q\bar{q}$ process to be 83\% for the COMPASS
experiment. This value is consistent with pQCD results shown in
Fig.~\ref{fig3_compass}. Applying this relative fraction of two pQCD
processes, the NNLO result of $\nu$ could be reasonably well
described, as shown in Fig.~\ref{fig4_compass}, with the acoplanarity
parameter $ \langle \cos 2\phi_1 \rangle$, set at 0.94. The predicted
$q_T$ distribution of the L-T violation $1-\lambda-2\nu$ from the NNLO
pQCD could be then nicely described as well.

Overall our studies show that salient features of $q_T/Q$ scaling and
$x_F$ dependency for the $\lambda$, $\nu$, $\mu$ parameters of
fixed-target D-Y experiments evaluated by NLO pQCD as well as the L-T
violation $1-\lambda-2\nu$ from the NNLO pQCD can be nicely understood
using the geometric picture.

\section{Summary and Conclusion}
\label{sec:conclusion}

We have presented a comparison of the measurements of the angular
parameters $\lambda$, $\mu$, $\nu$ and $1-\lambda-2\nu$ of the D-Y
process from the fixed-target experiments with the corresponding
results from the NLO and NNLO pQCD calculations. Qualitatively the
transverse momentum ($q_T$) dependence of $\lambda$, $\mu$ and $\nu$
in the data could be described by pQCD. The difference between NLO and
NNLO results becomes visible at large $q_T$. The L-T violation part
$1-\lambda-2\nu$ remains zero in the NLO pQCD calculation and turns
positive in NNLO pQCD. It is contrary to the measured negative values
in the pion-induced D-Y experiments NA10 and E615. Data quality, nonperturbative effects such as Boer-Mulders function at low $q_T$ and higher-order perturbative QCD at large $q_T$ might account for the discrepancy.


From the NLO pQCD calculation, we then present the predictions of the
angular parameters as a function of $q_T$ in several $Q$ and $x_F$
bins for the ongoing COMPASS and SeaQuest experiments. The $\lambda$
and $\nu$ show some mild dependence on $Q$ and a weak $x_F$
dependence, while $\mu$ exhibits a pronounced dependence on $x_F$. For
different $x_F$-values, $\lambda$ and $\nu$ are predicted to
approximately scale with $q_T/Q$.

The $x_F$ dependence of the angular parameters is well described by
the geometric picture. In particular, the weak rapidity dependencies
of the $\lambda$ and $\nu$, and the pronounced rapidity dependency for
$\mu$ can be explained by the absence or presence of
rapidity-dependent cancellation effects. The occurrence of
acoplanarity between the quark plane and the hadron plane ($\phi_1
\neq 0$), for the pQCD processes beyond NLO leads to a violation of
the $L-T$ relation. The predicted positive value of $1-\lambda-2\nu$,
or $A_0>A_2$ when $\phi_1$ is nonzero, is consistent with the NNLO
pQCD results.

The resummation effect of soft-gluon emission is not taken into
account in this work. In the geometric approach, summing over multiple
gluon emissions by a single quark line is equivalent to an emission of
a single gluon. Therefore, as long as the resummation is only
performed for a single quark, the L-T relation will still be
satisfied, as shown in Ref.~\cite{berger}. For a comprehensive pQCD
calculation, the resummation effect should be included, especially in
the small $q_T$ region. We leave it for future investigation.

The NLO and NNLO pQCD calculations should provide a good benchmark for
understanding the experimental data of lepton angular distributions of
fixed-target D-Y experiments. It is interesting to see many salient
features present in pQCD results can be readily understood by the
geometric picture. This intuitive approach could offer some useful
insights on the origins of many interesting characteristics of the
lepton angular distributions in the forthcoming new precision data
from the COMPASS and SeaQuest experiments. Any deviation from the pQCD
results on the L-T violation as well as the $\nu$ parameter would
indicate the presence of nonperturbative effects such as the
Boer-Mulders functions. Finally we emphasize the importance of
measuring the angular parameters in the D-Y process, which provides a
powerful tool to explore the reaction mechanism and parton
distributions potentially more sensitively than the D-Y cross sections
alone. The measurement of the $q_T$ distributions of $\mu$ parameters
with $x_F$ dependence is suggested, and the pQCD effect should be
included in the extraction of nonperturbative Boer-Mulders effect
from the data of $\nu$.

\section*{Acknowledgments}
\label{sec:acknowledgments}

This work was supported in part by the U.S. National Science
Foundation and the Ministry of Science and Technology of Taiwan. It
was also supported in part by the U.S. Department of Energy, Office of
Science, Office of Nuclear Physics under Contract No. DE-AC05-060R23177.



\begin{thebibliography}{40}

\bibitem{drell} S.~D.~Drell and T.~M.~Yan, Phys. Rev. Lett. {\bf 25}, 316
   (1970); Ann. Phys. (N.Y.) {\bf 66}, 578 (1971).

\bibitem{peng14} J.~C.~Peng and J.~W.~Qiu, Prog. Part. Nucl. Phys.  {\bf
  76}, 43 (2014).

\bibitem{falciano86} S.~Falciano {\em et al.}
  (NA10 Collaboration), Z. Phys. C {\bf 31}, 513 (1986); M.~Guanziroli
  {\em et al.}, Z. Phys. C {\bf 37}, 545 (1988).

\bibitem{conway} J.~S.~Conway {\em et al.} (E615 Collaboration), Phys.
  Rev. D {\bf 39}, 92 (1989); J.~G.~Heinrich {\em et al.}, Phys. Rev.
  D {\bf 44}, 1909 (1991).

\bibitem{dutta2013} W.~C.~Chang and D.~Dutta, Int. J. Mod. Phys. E {\bf
  22}, 1330020 (2013).

\bibitem{NA51} 
  A.~Baldit {\it et al.} (NA51 Collaboration),
  Phys.\ Lett.\ B {\bf 332}, 244 (1994).

\bibitem{e866} R.~S.~Towell {\em et al.} (E866/NuSea Collaboration),
  Phys.\ Rev.\ D {\bf 64}, 052002 (2001).

\bibitem{chang14} W.~C.~Chang and J.~C.~Peng, Prog. Part. Nucl. Phys.
{\bf 79}, 95 (2014).

\bibitem{Bacchetta:2017gcc}
  A.~Bacchetta, F.~Delcarro, C.~Pisano, M.~Radici and A.~Signori,
  J. High Energy Phys. {\bf 06} (2017) 081.

\bibitem{boer99} D. Boer, Phys. Rev. D {\bf 60}, 014012 (1999).


\bibitem{compass} M.~Aghasyan {\em et al.} (COMPASS Collaboration),
Phys. Rev. Lett. {\bf 119}, 112002 (2017).

\bibitem{lam78} C.~S.~Lam and W.~K.~Tung, Phys. Rev. D {\bf 18}, 2447
  (1978).

\bibitem{lam80} C.~S.~Lam and W.~K.~Tung, Phys. Rev. D {\bf 21}, 2712
  (1980).

\bibitem{Brandenburg:1993cj}
  A.~Brandenburg, O.~Nachtmann and E.~Mirkes,
  Z.\ Phys.\ C {\bf 60}, 697 (1993).

\bibitem{zhu} L.~Y.~Zhu {\em et al.} (Fermilab E866 Collaboration),
    Phys. Rev. Lett. {\bf 99}, 082301 (2007);
    Phys. Rev. Lett. {\bf 102}, 182001 (2009).

\bibitem{cdf}
  T.~Aaltonen {\it et al.} (CDF Collaboration),
  Phys.\ Rev.\ Lett.\  {\bf 106}, 241801 (2011).

\bibitem{cms} V.~Khachatryan {\em et al.} (CMS Collaboration), Phys.
Lett. B {\bf 750}, 154 (2015).

\bibitem{atlas} G. Aad {\em et al.} (ATLAS Collaboration),
J. High Energy Phys. {\bf 08} (2016) 159.

\bibitem{Gauld:2017tww} R.~Gauld, A.~Gehrmann-De Ridder, T.~Gehrmann,
  E.~W.~N.~Glover and A.~Huss,
  J. High Energy Phys. {\bf 11}, 003 (2017) 003.

\bibitem{Lambertsen:2016wgj} M.~Lambertsen and W.~Vogelsang,
  Phys.\ Rev.\ D {\bf 93}, 114013 (2016).

\bibitem{peng16} J.~C.~Peng, W.~C.~Chang, R.~E.~McClellan, and
  O.~Teryaev, Phys. Lett. B {\bf 758}, 384 (2016).

\bibitem{chang17} W.~C.~Chang, R.~E.~McClellan, J.~C.~Peng, and
  O.~Teryaev, Phys. Rev. D {\bf 96}, 054020 (2017).

\bibitem{peng18} 
  J.~C.~Peng, D.~Boer, W.~C.~Chang, R.~E.~McClellan and O.~Teryaev,
 Phys. Lett. B {\bf 789}, 356 (2019).

\bibitem{COMPASSII} COMPASS-II Proposal, Report No. CERN-SPSC-2010,
  \url{http://cds.cern.ch/record/1265628}.

\bibitem{E906} P.~E.~Reimer (Fermilab SeaQuest Collaboration),
  J. Phys. Conf. Ser. {\bf 295}, 012011 (2011).

\bibitem{DY_nlo} G.~Altarelli, R.~K.~Ellis and G.~Martinelli,
  Nucl.\ Phys. {\bf B157}, 461 (1979); J.~Kubar-Andre and
  F.~E.~Paige, Phys.\ Rev.\ D {\bf 19}, 221 (1979); K.~Harada,
  T.~Kaneko and N.~Sakai,
  Nucl.\ Phys.\ {\bf B155}, 169 (1979); {\bf B165}, 545(E) (1980).

\bibitem{DY_nnlo}
  R.~K.~Ellis, G.~Martinelli and R.~Petronzio,
  Nucl.\ Phys.\ {\bf B211}, 106 (1983); R.~J.~Gonsalves,
  J.~Pawlowski and C.~F.~Wai,
  Phys.\ Rev.\ D {\bf 40}, 2245 (1989); P.~B.~Arnold and M.~H.~Reno,
  Nucl.\ Phys.\ {\bf B319}, 37 (1989); {\bf B330}, 284(E) (1990).

\bibitem{DYNNLO} S.~Catani, L.~Cieri, G.~Ferrera, D.~de Florian, and
  M.~Grazzini, Phys. Rev. Lett. {\bf 103}, 082001 (2009); S.~Catani
  and M.~Grazzini, Phys. Rev. Lett. {\bf 98}, 222002 (2007).

\bibitem{FEWZ}
  K.~Melnikov and F.~Petriello,
  Phys.\ Rev.\ D {\bf 74}, 114017 (2006).

\bibitem{boer06} D.~Boer and W.~Vogelsang, Phys. Rev. D {\bf 74},
014004 (2006).

\bibitem{berger} E.~L.~Berger, J.~W.~Qiu, and R.~A.~Rodriguez-Pedraza,
  Phys. Lett. B {\bf 656}, 74 (2007); Phys. Rev. D {\bf 76}, 074006
  (2007).

\bibitem{DYNNLO_Web} DYNNLO v1.5,
  \url{http://theory.fi.infn.it/grazzini/dy.html}.

\bibitem{LHAPDF6}
  A.~Buckley, J.~Ferrando, S.~Lloyd, K.~Nordström, B.~Page, M.~Rüfenacht, M.~Schönherr and G.~Watt,
  Eur.\ Phys.\ J.\ C {\bf 75}, 132 (2015).

\bibitem{PDFsets} Official LHAPDF6 PDF sets,
  \url{https://lhapdf.hepforge.org/pdfsets.html}.

\bibitem{cs} J.~C.~Collins and D.~E.~Soper, Phys. Rev. D {\bf 16}, 2219
  (1977).

\bibitem{Zhang:2008nu}
  B.~Zhang, Z.~Lu, B.~Q.~Ma and I.~Schmidt,
  Phys.\ Rev.\ D {\bf 77}, 054011 (2008);
  Z.~Lu and I.~Schmidt,
  Phys.\ Rev.\ D {\bf 81} 034023 (2010);
X.~Wang, W.~Mao and Z.~Lu,
  Eur.\ Phys.\ J.\ C {\bf 78}, 643 (2018).

\bibitem{Barone:2009hw}
  V.~Barone, S.~Melis and A.~Prokudin,
  Phys.\ Rev.\ D {\bf 81}, 114026 (2010);
  V.~Barone, S.~Melis and A.~Prokudin,
  Phys.\ Rev.\ D {\bf 82}, 114025 (2010).

\bibitem{transversity2014} J.~C.~Peng, 
  EPJ Web Conf.\  {\bf 85}, 01009 (2015).

\bibitem{e1039} Fermilab Proposal 1039,
  Report. No. FERMILAB-PROPOSAL-103,
  \url{http://inspirehep.net/record/1309534}.
  
\bibitem{collins} J.~C.~Collins, Phys. Rev. Lett. {\bf 42}, 291 (1979).

\bibitem{thews} R.~L.~Thews, Phys. Rev. Lett. {\bf 43}, 987 (1979).

\bibitem{lindfors} J.~Lindfors, Phys. Scr. {\bf 20}, 19 (1979).

\end{thebibliography}
\end{document}